\documentclass[nofootinbib,amsmath,amssymb,aps,twocolumn,prd,preprintnumbers]{revtex4-2}
\usepackage{amsmath,latexsym,mathtools}
\usepackage[font=small,labelfont=bf,justification=justified,format=plain]{caption}
\usepackage{ragged2e}

\usepackage[x11names,dvipsnames,table]{xcolor} 
\usepackage[normalem]{ulem}
\usepackage{dcolumn}
\usepackage{subfig}

\usepackage[pscoord]{eso-pic}

\usepackage{makecell}
\usepackage{booktabs} 

\usepackage{etoolbox}
\usepackage{amsmath,latexsym,mathtools}
\usepackage{comment}
\usepackage{footnpag}
\usepackage{float} 
\usepackage{xcolor}    
\usepackage[multiple]{footnotehyper}
\usepackage{chngcntr}
\counterwithout{footnote}{section} 

\usepackage{slashed}
\usepackage{braket}
\usepackage{amssymb}
\usepackage{amsfonts}
\usepackage{float}
\usepackage[utf8]{inputenc}
\usepackage{amsmath}
\usepackage{physics}
\usepackage{graphicx} 
\usepackage{dcolumn} 
\usepackage{bm} 

\usepackage{subcaption}
\usepackage{changepage}
\usepackage{xcolor}
\usepackage{float} 

\usepackage[colorlinks]{hyperref}
\usepackage[thinc]{esdiff}

\AtBeginDocument{\hypersetup{
	colorlinks=true,
	linkcolor=black!40!blue,
	citecolor=black!40!blue,
	filecolor=black,
	urlcolor=black!45!blue,
}}

\usepackage{comment}

\begin{document}

\title{Multifield theories invariant under transverse diffeomorphisms: The mixed regime case}

\author{Antonio L. Maroto}
\email{maroto@ucm.es}
\affiliation{Departamento de Física Teórica and Instituto de Física de Partículas y del Cosmos (IPARCOS-UCM), Universidad Complutense de Madrid, 28040, Madrid, Spain}

\author{Prado Martín-Moruno}
\email{pradomm@ucm.es}
\affiliation{Departamento de Física Teórica and Instituto de Física de Partículas y del Cosmos (IPARCOS-UCM), Universidad Complutense de Madrid, 28040, Madrid, Spain}

\author{Diego Tessainer}
\email{dtessain@ucm.es}
\affiliation{Departamento de Física Teórica and Instituto de Física de Partículas y del Cosmos (IPARCOS-UCM), Universidad Complutense de Madrid, 28040, Madrid, Spain}


\date{\today}
\begin{abstract}
We study theories breaking diffeomorphism (Diff) invariance down to the subgroup of transverse diffeomorphisms (TDiff), consisting of multiple scalar fields in a cosmological background. In particular, we focus on models involving a field dominated by its kinetic term and a field dominated by its potential, coupled to gravity through power-law functions of the metric determinant. The Diff symmetry breaking results in the individual energy-momentum tensors not being conserved, although the total conservation-law is satisfied. Consequently, an energy exchange takes place between the fields, acting as an effective interaction between them. With this in mind, we consider the covariantized approach to describe the theory in a Diff invariant way but with an additional field, and discuss the phenomenological consequences of these models when it comes to the study of the dark sector.

\end{abstract}

\preprint{IPARCOS-UCM-25-043}
\maketitle

\section{Introduction}

As is well known, General Relativity (GR) constitutes the current best description of gravity. However, there are some issues both at the observational and theoretical level that may hint at the necessity to consider extensions or modifications of this theory. Observational data indicate that our universe currently exhibits an accelerated expansion \cite{Riess_1998}, which in the standard cosmological model is considered to be driven by a dark energy component depicted by a cosmological constant. Nevertheless, the expected size obtained from quantum field theory for this constant is much higher than the measured one, this issue being commonly referred to as the vacuum-energy problem. In addition,
there exists a well-known tension between local and cosmological measurements of the Hubble parameter $H_0$ \cite{Riess:2019cxk,DiValentino:2020zio,Abdalla:2022yfr}. 
Some alternative dark energy models involve a dynamical scalar field, such as \textit{quintessence} \cite{Tsujikawa:2013fta}, \textit{k-essence} \cite{Armendariz-Picon:2000ulo}, or \textit{phantom} models \cite{DiValentino:2020naf}. In this latter case, the energy density would increase with the expansion. These models have been deeply analyzed, with it being shown that tensions in the measurements of cosmological parameters could particularly be alleviated when considering phantom dark energy or dark matter-dark energy interactions \cite{Heisenberg:2022lob,DiValentino:2021izs}. However, these interactions are primarily phenomenological and lack a fundamental motivation. In particular, the combined analysis of DESI (BAO) $+$ CMB $+$ SNe data favors dynamical dark energy crossing the phantom line in the past \cite{DESI:2025zgx}.

As a consequence of these issues, modifications of gravity at cosmological scales have been considered, commonly inspired by several extensions of GR (for a review, see \cite{Clifton:2011jh}). One of the possible modifications that has been analyzed in detail in recent years is the breaking of diffeomorphism (Diff) invariance of the theory. Diffeomorphism invariance (i.e., physical invariance under general coordinate transformations) is the fundamental symmetry of GR and there are some alternative theories that restrict it. Particularly, in unimodular gravity \cite{einsteinUG,PhysRevD.40.1048,Carballo-Rubio_2022}, the metric determinant $g$ is taken to be a non dynamical field fixed to the value $g=1$ and Diff invariance is broken down to the subgroup of transverse diffeomorphisms (TDiff) and Weyl rescalings. As a matter of fact, unimodular gravity theories have been shown to provide a solution to the vacuum-energy problem \cite{Ellis:2010uc}.

Theories invariant under transverse diffeomorphisms have been gaining popularity over the recent years and they are of special interest for describing the cosmological dark sector. In these theories the full Diff invariance is broken down to general coordinate transformations whose Jacobian $\mathcal{J}$ is equal to one, which results in previously tensor densities becoming true tensors under TDiff transformations. TDiff theories that break the symmetry through the gravitational sector have been studied in \cite{Toni1:2023btf,Toni2:2024vqk}, while theories breaking the symmetry through the matter sector have been investigated in \cite{Maroto:only,Dario1:2023cor,Dario2:2024tdv,DAWEI:2023hkx,dpm:2024ewm,JdeCruzPerez:2025ytd}. 
In particular, \cite{Maroto:only} focuses on the case of a single TDiff scalar field theory in a cosmological context and \cite{Dario1:2023cor} is devoted to the study of a general TDiff scalar field theory without assuming any specific background, while \cite{DAWEI:2023hkx} and \cite{dpm:2024ewm} revolve around the particular applications of TDiff theories when it comes to the dark sector from a more phenomenological point of view. On the other hand, in reference \cite{Dario2:2024tdv} an equivalent way to describe a TDiff invariant theory as a Diff theory with an extra field is developed and in reference \cite{JdeCruzPerez:2025ytd} this approach is applied to show that $\Lambda$CDM can be written as a particular TDiff invariant matter model. It should be noted that these theories are of particular interest for applications to the dark sector, as dark fields could have different symmetries than the visible sector. The analysis of multi-field TDiff theories was performed in \cite{dpm:2024ewm}, where the cosmology of shift-symmetric models at the background level was studied. In particular, even for fields without explicit interactions in the Lagrangian, the Diff breaking induces an effective coupling with interesting phenomenological consequences.

In this work we will focus on the extension of the multi-field models beyond the shift symmetric case. In particular, we will focus for the first time on the case in which the matter sector is constituted by two free scalar fields; the dynamics of one of them being entirely dominated by its kinetic term and the dynamics of the other by its potential term. The main objective of this work is to explore this specific model and its dynamics from a theoretical point of view, and present a particular application when it comes to the dark sector. In a single-field scenario, 
the second field would behave like a cosmological constant in the potential domination regime. However, the presence of the other field in its kinetic domination regime will entail an effective interaction between both fields induced by the Diff symmetry breaking, and it will be reflected in the conservation law for the total EMT. This will result in the possibility of the potentially driven field depicting a dynamical dark energy component, which could exhibit a wide range of possible behaviors 
when the other field dominates.

The paper is organized as follows. In section \ref{secII} we review the foundations of single-field TDiff theories in
the TDiff formalism and the solutions that can be obtained in the respective potential and kinetic domination regimes. Section \ref{secIII} is dedicated to discuss the covariantized approach and its relation with the TDiff framework, presenting the equivalences between both of them within the context of scalar field theories. Section \ref{secIV} contains the main body of the paper. We shall consider a TDiff two-field theory and introduce the mixed regime case, which involves a kinetically driven field and a potentially driven field. Focusing on power-law coupling functions and following the covariantized formalism, we investigate the single field domination regimes in section \ref{secIVA}, analyze the direction of the energy exchange in section \ref{secIVB}, and examine the dark sector phenomenology in section \ref{secIVC}.
In section \ref{secV} we will discuss the conclusions. Finally, in appendix \ref{appendix}, we further discuss the relation between the TDiff and the covariantized formalisms, exploring the model in the TDiff framework and discussing the advantages of working with the covariantized approach.

\section{TDiff formalism}\label{secII}
In this section we will review the results obtained for single TDiff invariant scalar field theories using the TDiff approach, but let us first briefly introduce transverse diffeomorphisms \cite{Maroto:only}. If we consider a general infinitesimal coordinate transformation $x^\mu\mapsto \tilde x^\mu=x^\mu+\xi^\mu(x)$, where $\xi^\mu(x)$ is the vector field driving the transformation, the variation of the metric tensor $g_{\mu\nu}(x)$ will read \cite{Maroto:only}
\begin{equation}
    \delta g_{\mu\nu}=\mathfrak{L}_\xi(g_{\mu\nu})=-\nabla_\nu\xi_\mu-\nabla_\mu\xi_\nu,
    \label{metric_variation}
\end{equation}
where $\mathfrak{L}_\xi(g_{\mu\nu})$ denotes the Lie derivative of the metric tensor. Thus, the metric determinant $g\equiv|\mathrm{det}(g_{\mu\nu})|$ will transform the following way \footnote{Since the metric determinant does not change under orientation changes,
the integrals will change sign under orientation non-preseverving transverse diffeomorphisms, very much as in Diff theories.}:
\begin{equation}
    \delta g=gg^{\mu\nu}\delta g_{\mu\nu}=-2g\nabla_\mu\xi^\mu.
    \label{determinant_variation}
\end{equation}
If we then consider a matter action given by a Diff-invariant Lagrangian density $\mathcal{L}(g_{\mu\nu}(x),\phi(x),\partial_\mu(x))$ involving a scalar field $\phi(x)$, i.e.,
\begin{equation}
    S_\mathrm{mat}[g_{\mu\nu},\phi]=\int\dd^4xf(g)\mathcal{L}(g_{\mu\nu}(x),\phi(x),\partial_\mu\phi(x)),
    \label{matter_action_first}
\end{equation}
with an arbitrary coupling function of $g$ given by $f(g)$, we can compute the variation of the matter action under $\xi(x)$. This yields 
\begin{equation}
    \delta_\xi S_\mathrm{mat}=\int \dd^4 x\partial_\mu\xi^\mu[f(g)-2gf'(g)]\mathcal{L},
    \label{variation_action}
\end{equation}
where a prime denotes derivative with respect to the argument of the function. It can be noted that $\delta_\xi S_\mathrm{mat}=0$ either when $f(g)-2gf'(g)=0$, which is equivalent to Diff invariance, or when $\partial_\mu\xi^\mu=0$, which is the transversality condition required to be satisfied by transverse diffeomorphisms. It should be emphasized that in the second case the coupling function $f(g)$ is kept arbitrary.

Let us now consider a general matter action involving a single scalar field:
\begin{equation}
    S_\mathrm{mat}=\int\dd^4x\left[\frac{1}{2}f_K(g)\partial_\mu\phi\partial^\mu\phi-f_V(g)V(\phi)\right],
    \label{1scalar_field_action}
\end{equation}
which clearly breaks Diff invariance down to TDiff invariance, and where $f_K(g)$ and $f_V(g)$ denote the coupling functions of the kinetic and potential terms, respectively, which are chosen to be positive-valued in order to avoid instabilities \cite{Maroto:only,Dario1:2023cor}. The general action will then read
\begin{equation}
    S=S_\mathrm{EH}[g_{\mu\nu}]+S_\mathrm{mat}[g_{\mu\nu},\phi],
    \label{general_action}
\end{equation}
where $S_\mathrm{EH}$ denotes the usual Einstein-Hilbert action
\begin{equation}
   S_\mathrm{EH}=-\frac{1}{16\pi G}\int\dd^4 x\sqrt{g}R,
    \label{actionEH}
\end{equation}
since we are not breaking Diff invariance through the gravity sector. The equations of motion for the gravitational field will thus be the usual Einstein equations \cite{Maroto:only}:
\begin{equation}
    G_{\mu\nu}=8\pi GT_{\mu\nu},
    \label{Einstein_equations}
\end{equation}
where $G_{\mu\nu}$ is the Einstein tensor, $G_{\mu\nu}=R_{\mu\nu}-\frac{1}{2}Rg_{\mu\nu}$, and $T_{\mu\nu}$ is the EMT, which is calculated from its usual definition
\begin{equation}
    T_{\mu\nu}=\frac{2}{\sqrt{g}}\frac{\delta S_\mathrm{mat}}{\delta g^{\mu\nu}}. 
    \label{EMT_definition}
\end{equation}
Taking into account (\ref{1scalar_field_action}), the EMT for the TDiff field reads
\begin{eqnarray}
    T_{\mu\nu}&=&\frac{f_K(g)}{\sqrt{g}}\left[\partial_\mu\phi\partial_\nu\phi-F_K(g)g_{\mu\nu}\Box\phi\right]\nonumber \\ &+&2\frac{f_V(g)}{\sqrt{g}}F_V(g)g_{\mu\nu}V(\phi),
    \label{emt_gen_1field}
\end{eqnarray}
where $F_i(g)\equiv\dd\ln f_i(g)/\dd\ln g$. It is worth noting that, since the field equations are the usual Einstein equations  but with a modified EMT, some properties of GR are maintained. For instance, the Cauchy problem is the same as in GR, and it is thus well posed. Similarly, we will also recover the Newtonian limit because the field equations are the same as in GR (the only difference is the modified EMT). When it comes to the scalar field, its EoM can be written as
\begin{equation}
\partial_\mu[f_K(g)\partial^\mu\phi]+f_V(g) V'(\phi)=0.
    \label{EoM_scalar_field}
\end{equation}
It is also worth mentioning that the Equivalence Principle is not violated in models with $f_K(g)=f_V(g)$ so that, also in the TDiff case, particles propagate along geodesics (see further details in \cite{Maroto:only}).

With regards to the conservation of the EMT, it is known that in GR this is an automatic consequence of Diff invariance through Noether theorem. However, in TDiff theories the conservation of this quantity does not trivially follow. In addition, in TDiff we have one less gauge freedom for choosing the coordinates (recall the transversality condition $\partial_\mu\xi^\mu=0$). Nevertheless, since Einstein equations hold and $\nabla_\mu G^{\mu\nu}=0$ {\color{black} due to Bianchi identities}, one has that $\nabla_\mu T^{\mu\nu}=0$ on solutions of Einstein equations. This consistency condition will allow us to fix the extra degree of freedom related to having less gauge symmetry, i.e., it is a physical constraint on the metric \cite{Maroto:only,Dario1:2023cor}.

In particular, this work will be focused on the spatially flat Friedmann-Lemaître-Robertson-Walker (FLRW) spacetime, defined by \cite{E_Alvarez_Faedo.76.064013}
\begin{equation}
    \dd s^2=b(\tau)^2\dd\tau^2-a(\tau)^2\mathbf{dx}^2,
    \label{flat_FLRW_spacetime}
\end{equation}
with $g(\tau)=b(\tau)^2a(\tau)^6$. Notice that unlike in GR, we cannot generally fix the shift function $b(\tau)$ since time reparametrizations are not TDiff transformations. However, the constraint that follows from the EMT conservation law will allow us to obtain a relation between $b(\tau)$ and the scale factor $a(\tau)$.

For the purpose of this work, we will briefly recap the results obtained for single scalar fields in the potential and kinetic domination regimes, which will be the basis to the multi-field model we will later present. Our only assumption will be that $\partial_\mu\phi$ is a timelike vector \cite{Dario1:2023cor}, so that we can express the EMT as that of a perfect fluid:
\begin{equation}
    T_{\mu\nu}=(\rho+p)u_\mu u_\nu-pg_{\mu\nu},
    \label{EMT_perfect_fluid}
\end{equation}
where $\rho$ and $p$ are the energy density and pressure, respectively; and $u_\mu=\partial_\mu\phi/\mathcal{N}$ is the four-velocity, {\color{black} with $\mathcal{N\equiv\sqrt{\partial_\mu\phi\partial^\mu\phi}}$ a normalization factor}. It is also worth remarking that, for the rest of the work we will assume that the background fields considered are homogeneous, i.e., $\phi=\phi(\tau)$.

\subsection{Potential domination regime}\label{secIIA}
In the case in which the potential term is dominant over the kinetic one, the EoM for the scalar field $\phi$ simply yields \cite{Maroto:only}
\begin{equation}
    V'(\phi)=0,
    \label{EoM_phi_potential}
\end{equation}
which indicates that $\phi=C_\phi$ is a constant and equal to the value that optimizes the potential. On the other hand, the conservation law for the EMT translates into 
\begin{equation}
    \frac{\dd }{\dd\tau}\left[f_V'(g)\sqrt{g}\right]=0,
    \label{constr_1pot}
\end{equation}
which is satisfied either when $f_V(g)=A+B\sqrt{g}$, with $A,B=\mathrm{const.}$ $\forall g$ (this is equivalent to the Diff case {\color{black} when $A=0$} see \cite{Maroto:only,Dario2:2024tdv}) or when $g=\mathrm{const.}$. The latter corresponds to the TDiff case and implies that $b\propto a^{-3}$. The energy density and pressure can be written using \eqref{EMT_perfect_fluid} and recalling that $\rho=T^0{}_0$ and $p=-T^i{}_j\delta^j{}_i/3$:
\begin{eqnarray}
    \rho&=&2\frac{f_V(g)}{\sqrt{g}}F_V(g)V(\phi),
    \label{rho_1pot} \\
    p&=&-\rho.
    \label{p_1pot}
\end{eqnarray}
So, $w=p/\rho=-1$,
which indicate that the behavior of a single TDiff scalar field dominated by its potential term will be that of a cosmological constant \cite{Maroto:only,Dario2:2024tdv}.

\subsection{Kinetic domination regime}\label{secIIB}
Let us now focus our attention on the case in which the kinetic term of the scalar field is dominant. The EoM for $\phi$ will yield \cite{Maroto:only}
\begin{equation}
    \phi'(\tau)=\frac{C_\phi}{L(\tau)},
    \label{EoM_1kin_integral}
\end{equation}
where $L(\tau)\equiv f(g(\tau))/b(\tau)^2$ and $C_\phi=\mathrm{const.}$ Using \eqref{EMT_perfect_fluid} and the EoM, the energy density and pressure read 
\begin{eqnarray}
    \rho&=&\frac{f_K(g)}{b^2\sqrt{g}}[1-F_K(g)]\frac{C_\phi^2}{L(\tau)^2},
    \label{rho1kin} \\
    p&=&w\rho,
    \label{p1kin}
\end{eqnarray}
where
\begin{equation}
   w=\frac{F_K(g)}{1-F_K(g)}.
    \label{w_1kin_general}
\end{equation}
This indicates that the equation of state (EoS) parameter for a free kinetic TDiff scalar field will generally depend on $g$, and thus on time $\tau$. When it comes to the conservation law for the EMT, we can obtain the constraint using the zeroth component of $\nabla_\nu T^{\mu\nu}=0$, which reads:
\begin{equation}
    \rho'(\tau)+3\frac{a'(\tau)}{a(\tau)}[\rho(\tau)+p(\tau)]=0.
    \label{consT001kin}
\end{equation}
Substituting \eqref{EoM_1kin_integral}, \eqref{rho1kin} and \eqref{p1kin}, one can obtain the geometrical constraint required for the local conservation of the EMT in a compact way \cite{Maroto:only}:
\begin{equation}
    \frac{g}{f_K(g)}[1-2F_K(g)]=\mathcal{C}a^6,
    \label{constr_1kin_general}
\end{equation}
where $\mathcal{C}$ is a constant. This will allow us to fix the additional degree of freedom of the metric $b(\tau)$ in terms of the scale factor. 

\section{COVARIANTIZED FORMALISM}\label{secIII}
After having introduced the single-field results in the TDiff formalism, we will quickly recap these in the covariantized approach developed in  \cite{Dario2:2024tdv}, which introduces a Stueckelberg field. Reference \cite{Dario2:2024tdv} follows Henneaux and Teitelboim \cite{HENNEAUX1989195} in a similar procedure to the Stueckelberg approach in gauge field theories \cite{STUECKELBERG1}, and introduces a Diff scalar density $\Bar{\mu}$ which transforms as $\sqrt{g}$ under general coordinate transformations. This way, we can write our covariantized action as \cite{Dario2:2024tdv}
\begin{equation}
    S_\mathrm{mat}^\mathrm{cov}=\int \dd^4 x\sqrt{g}\left[\frac{\Bar{\mu}}{\sqrt{g}}f\left(\frac{g}{\Bar{\mu}^2}\right)\right]\mathcal{L}(g_{\mu\nu},\phi,\partial_\mu\phi),
    \label{S_cov_general}
\end{equation}
which is equivalent to \eqref{matter_action_first} in the case when $\Bar{\mu}=1$, which will be denoted as the TDiff frame; and where the term in brackets is a Diff scalar. For the sake of simplicity, we will consider the following notation for the rest of the work:
\begin{eqnarray}
    Y&=&\frac{\Bar{\mu}}{\sqrt{g}},
    \label{notation_1} \\
    H(Y)&=& Yf(Y^{-2}).
    \label{notation_2}
\end{eqnarray}
Following \cite{Dario2:2024tdv}, the new Stueckelberg field is chosen to be a vector. The scalar density $\Bar{\mu}$ can be related to this new field through 
\begin{equation}
    \Bar{\mu}=\partial_\mu(\sqrt{g}A^\mu), 
    \label{mu_T}
\end{equation}
i.e., $Y=\nabla_\mu A^\mu$, which allows us to finally write our general matter action as
\begin{equation}
    S_\mathrm{mat}^\mathrm{cov}=\int \dd^4 x\sqrt{g}[H_K(Y)X-H_V(Y)V(\phi)],
    \label{S_cov_general_Y}
\end{equation}
which is equivalent to \eqref{matter_action_first} in the TDiff frame, and where 
\begin{equation}
    X=\frac{1}{2}\,\partial_\mu\phi\,\partial^{\mu}\phi,
    \label{X_kin}
\end{equation}
where $H_i(Y)>0$ for stability purposes. Notice that, thanks to the introduction of $A^\mu$, this new action is Diff-invariant.

In the covariantized approach, the EoM for the scalar field $\phi$ and the EoM for the vector field $A^\mu$ read
\begin{equation}
    \nabla_\nu[H_K(Y)\nabla^\mu\phi]+H_V(Y)V'(\phi)=0,
    \label{EoM_scalar_field_cov}
\end{equation}
and
\begin{equation}
    \partial_\nu[H_K'(Y)X-H_V'(Y)V(\phi)]=0,
    \label{EoM_Tmu_cov}
\end{equation}
respectively.
It should be noted that in the covariantized formalism the EMT conservation is automatically satisfied on solutions of the field EoM, but there is an EoM for the additional vector field. As it was shown in reference \cite{Dario2:2024tdv}, that EoM is equivalent to the constraint when {\color{black} going to the TDiff frame}. As we can see, the EoM for the Stueckelberg field $A^\mu$ \eqref{EoM_Tmu_cov} acts as a constraint on $Y=\partial_\mu A^\mu$, since $A^\mu$ only appears through its divergence. 
This implies that there is a symmetry under the following transformations:
\begin{equation}
    A^\mu\mapsto\tilde{A}^\mu=A^\mu+D^\mu,
\end{equation}
for any divergenceless vector field $D^\mu$. 
Consequently, we can always fix a gauge in which $A^0=0$ and thus $A^\mu$ does not propagate. Thus, the number of propagating degrees of freedom reduces to the
scalar field $\phi$, being the Stueckelberg field just a non-propagating auxiliary field. Lastly, the EMT in the covariantized approach can be computed from equations \eqref{EMT_definition} and (\ref{S_cov_general_Y}). This is \cite{Dario2:2024tdv}
\begin{eqnarray}
    T_{\mu\nu}&=&H_K(Y)\partial_\mu\phi\partial_\nu\phi-[H_K(Y)X-H_V(Y)V(\phi)]g_{\mu\nu} \nonumber \\
    &+&Y[H_K'(Y)X-H_V'(Y)V(\phi)]g_{\mu\nu}.
    \label{EMT_cov}
\end{eqnarray}
In principle,  the conclusions about the validity of the Equivalence Principle obtained in the TDiff formalism in \cite{Maroto:only} can also be translated to the  covariantized formalism, although the details will be presented elsewhere.  In any case,
this does not affect any of the conclusions of the present paper.
We will now recap the results for the potential and kinetic domination regimes in the covariantized approach.
\subsection{Potential domination regime}
In the case in which the action of the scalar field is dominated by its potential term, the EoM for $\phi$ reads
\begin{equation}
    H_V(Y)V'(\phi)=0,
    \label{EoM_phi_cov_pot}
\end{equation}
which is non-trivially satisfied when $V'(\phi)=0$, implying that $\phi$ takes a constant value which optimizes the potential, equivalently to the TDiff case. On the other hand, the EoM for the vector field $A^\mu$ yields
\begin{equation}
    \partial_\mu[H_V'(Y)V(\phi)]=0\implies H_V''(Y)\partial_\mu Y=0,
    \label{EoM_T^mu_cov_pot}
\end{equation}
The first possibility for \eqref{EoM_T^mu_cov_pot} to be satisfied is that $H_V''(Y)=0$ and thus $H_V(Y)=\alpha Y+\beta$, with $\alpha,\beta=\mathrm{const.}$ (which, {\color{black} when $\alpha=0$}, corresponds to the Diff case, see \cite{Dario2:2024tdv}). The second solution occurs when $\partial_\mu Y=0$, which implies that $Y=\mathrm{const.}$ and consequently translates into $g=\mathrm{const.}$ in the TDiff frame. For completion, we present the expressions for the energy density and pressure in this formalism. These are
\begin{eqnarray}
    \rho&=&V(\phi)[H_V(Y)-YH_V'(Y)],
    \label{rho1p_cov}\\
    p&=&-\rho,
    \label{p1p_cov}
\end{eqnarray}
which indicate that $\phi$ will depict a cosmological constant, {\color{black}as expected} (this can also be checked by looking at the specific EMT under potential domination).

\subsection{Kinetic domination regime}
When the action of the matter field is dominated by its kinetic term, the EoM for the vector field $A^\mu$ reads
\begin{equation}
    \partial_\mu[H_K'(Y)X]=0\implies H_K'(Y)X=\mathrm{const.}
    \label{EoM_phi_cov_1kin}
\end{equation}
On the other hand, the EoM for $\phi$ reads
\begin{equation}
    \nabla_\mu[H_K(Y)\partial^\mu\phi]=\nabla_\mu[H_K(Y)\mathcal{N}u^\mu]=0,
    \label{EoM_T^mu_cov}
\end{equation}
 which allows us to finally obtain \cite{Dario2:2024tdv}:
\begin{equation}
    \phi'(\tau)^2=\frac{C_\phi^2}{H_K(Y)^2a^6},
    \label{EoM_T^mu_1kin_cov}
\end{equation}
which, when substituted in \eqref{EoM_phi_cov_1kin} yields:
\begin{equation}
    -\frac{H_K'(Y)}{H_K(Y)^2}=\mathcal{C}a^6,
    \label{constr_1kin_cov}
\end{equation}
where $\mathcal{C}$ is a constant. It is straightforward to see that {\color{black} going to the TDiff frame in \eqref{constr_1kin_cov}} allows us to recover the constraint \eqref{constr_1kin_general} from the TDiff formalism. Lastly, we present the expressions for the energy density and the EoS parameter of   
$\phi$ when the kinetic term is dominant:
\begin{eqnarray}
    \rho&=&X[H_K(Y)+YH_K'(Y)],
    \label{rho1kin_cov}\\
w&=&\frac{H_K(Y)-YH_K'(Y)}{H_K(Y)+YH_K'(Y)},
    \label{rho2kin_cov}
\end{eqnarray}
which indicates that $w=w(Y)$ and, therefore, the EoS parameter will evolve with the expansion of the universe.
\section{Multi-field interacting model}\label{secIV}
In this section we will present the specific model that we shall discuss in the rest of the work, which involves two free TDiff homogeneous
scalar fields. The most general action involving two non-interacting fields to lowest order in derivatives is, in the TDiff formalism,
\begin{eqnarray}
        S_\mathrm{mat}&=&\int\dd^4x\left[\frac{1}{2}f_{K1}(g)\partial_\mu\phi_1\partial^\mu\phi_1-f_{V1}(g)V_1(\phi_1)+\right. \nonumber \\
        &&\left.+\frac{1}{2}f_{K2}(g)\partial_\mu\phi_2\partial^\mu\phi_2-f_{V2}(g)V_2(\phi_2)\right],
    \label{S_mixto_gen_TDiff}
\end{eqnarray}
and, in the covariantized approach,
\begin{eqnarray}
    S_\mathrm{mat}^\mathrm{cov}&=&\int \dd^4x \sqrt{g}\left[H_{K1}(Y)X_1-H_{V1}(Y)V_1(\phi_1)+\right. \nonumber  \\ 
    &&\left. +H_{K2}(Y)X_2-H_{V2}(Y)V_2(\phi_2)\right].
    \label{S_mixto_gen_cov}
\end{eqnarray}
However, we shall now consider some specific assumptions for the particular model of interest. In the first place,
we will assume that $\phi_2$ is dominated by its kinetic term. {\color{black} This is the exact case in} theories that are symmetric under shift transformations of $\phi_2$, i.e., $\phi_2\mapsto\phi_2+C$, with $C$ an arbitrary constant. In the second place,
we will consider that the dynamics of $\phi_1$ is dominated by its potential contribution during the matter and dark energy epochs and that its kinetic contribution will be negligible at such time (a deeper justification of this approximation can be found later on). We shall consider such a model to be in the mixed regime\footnote{For an analysis of multi-field TDiff theories at the background level involving two shift-symmetric scalar fields, see \cite{dpm:2024ewm}.}. Thus, the matter action of our model will read, in the TDiff approach,
\begin{eqnarray}
    S_\mathrm{mat}&=&\int\dd^4x\left[-f_1(g)V_1(\phi_1)+f_2(g)X_2\right],
    \label{S_model_TDiff}
\end{eqnarray}
and, in the covariantized formalism,
\begin{eqnarray}
    S_\mathrm{mat}^\mathrm{cov}&=&\int \dd^4x \sqrt{g}\left[-H_1(Y)V_1(\phi_1)+H_2(Y)X_2\right],
    \label{S_model_cov}
\end{eqnarray}
{\color{black} where from now on we will denote $f_1(g)\equiv f_{V_1}(g), f_2(g)\equiv f_{K_2}(g)$,} $H_1(Y)\equiv H_{V_1}(Y)$ and $H_2(Y)\equiv H_{K_2}(Y)$ for simplicity. For now we will proceed considering both approaches. However, after having introduced the main consequences of breaking Diff symmetry we will focus on the covariantized approach, since it is more compact to work with for the present study.

As it is shown in the action, the fields are non interacting with each other, so the total EMT will simply read 
\begin{eqnarray}
    T_{\mu\nu}=T_{\mu\nu}^{(1)}+T_{\mu\nu}^{(2)},
      \label{EMT_total_modelo}
\end{eqnarray}
where
\begin{eqnarray}
T_{\mu\nu}^{(i)}=(\rho_i+p_i)u_\mu u_\nu-p_ig_{\mu\nu},
\end{eqnarray}
and
both components share a common four-velocity $u_\mu$ because they are homogeneous. Similarly, the respective energy density and pressure of each field will be given by the expressions presented in the previous sections. As we have discussed,
the total EMT must be conserved, i.e.,
\begin{equation}
    \nabla_\mu T^{\mu\nu}=\nabla_\mu T^{(1)\mu\nu}+\nabla_\mu T^{(2)\mu\nu}=0,
    \label{cons_EMT_total}
\end{equation}
which does not imply the conservation of the individual EMTs. In the FLRW background this can be expressed as
\begin{eqnarray}
    \rho_1'(\tau)+3\frac{a'(\tau)}{a(\tau)}[\rho_1(\tau)+p_1(\tau)]&=&Q(\tau),
    \label{Q_field1} \\
    \rho_2'(\tau)+3\frac{a'(\tau)}{a(\tau)}[\rho_2(\tau)+p_2(\tau)]&=&-Q(\tau),
    \label{Q_field2}
\end{eqnarray}
where $Q(\tau)$ will be referred to as the interacting kernel. The conservation law \eqref{cons_EMT_total} combined with the EoM of the fields will allow us to obtain a constraint on the metric. {The EoM of each field as well as its EMT have been presented in the previous section, since there is no interaction term in the action.}
In particular, substituting \eqref{EoM_phi_potential} for $\phi_1$ and \eqref{EoM_1kin_integral} for $\phi_2$ in \eqref{EMT_definition}, considering equation \eqref{cons_EMT_total} and recalling that $g=b^2a^6$ yields
\begin{equation}
    \frac{\dd}{\dd\tau}[2V_1\sqrt{g}f_1'(g)]+\frac{1}{\sqrt{g}}\frac{\dd}{\dd\tau}\left[C_{\phi_2}^2\frac{g|2F_2(g)-1|}{a^6f_2(g)}\right]=0.
    \label{constr_mixto_general}
\end{equation}
Thus, the constraint takes the form of an ODE that will allow us to obtain $b$ in terms of $a$. It is worth noting that this expression is different from the constraints obtained for the single-field cases \eqref{constr_1pot} and \eqref{constr_1kin_general}, since it has contributions from both components. This means that the $b(a)$ relation will be altered by the presence of multiple fields, affecting the behavior of each EMT through their energy densities and pressure. More specifically, even if there is no interaction term in the Lagrangian, the constraint \eqref{constr_mixto_general} will affect both components, acting as source of an effective interaction between them. Particularly, this could lead to the field in the potential regime behaving like a dynamical dark energy component, with its behavior differing considerably from that of a cosmological constant.

We will now show how we can also work equivalently using the covariantized approach. In this case, as we previously discussed, the geometrical constraint can be derived from the EoM for the vector field $A^\mu$ (after taking the TDiff limit), which in this case read
\begin{equation}
    \partial_\nu[H_2'(Y)X-H_1'(Y)V(\phi)]=0,
\end{equation}
which implies
\begin{equation}
    H_2'(Y)X-H_1'(Y)V(\phi)=\mathrm{const.}
    \label{EoM_T^mu_mixto_general}
\end{equation}
If we then substitute \eqref{EoM_phi_cov_pot} and \eqref{EoM_T^mu_1kin_cov}  in this equation and go to the TDiff frame, i.e., $\Bar{\mu}\mapsto1$ ($Y\mapsto1/\sqrt{g}$, $H_i(Y)\mapsto f_i(g)/\sqrt{g}$), \eqref{EoM_T^mu_mixto_general} translates into an expression that is equivalent to the constraint \eqref{constr_mixto_general}. However, it should be noted that \eqref{EoM_T^mu_mixto_general} is a {\color{black} more manageable expression} than \eqref{constr_mixto_general}. 

\subsection*{Power-law coupling functions}
From now on we will analyze all the details and the phenomenology of the model using the covariantized approach, since the analysis is simpler and more compact\footnote{In appendix \ref{appendix} we discuss this choice in further detail and explore the model in the TDiff framework.}. 
For simplicity we focus on power-law functions $H_i(Y)=\lambda_iY^{\alpha_i}$ with positive $\lambda_i$, which are of particular interest. In fact, this choice has the advantage that, when moving back to the TDiff frame, a power-law in $H_i(Y)$ corresponds to a power-law in $f_i(g)$, with a different power \cite{Dario2:2024tdv}.  In this scenario, it is easy to obtain the expression of the EoS parameters (see \eqref{p_1pot} and \eqref{w_1kin_general}). These are:
\begin{equation}
    w_1=-1,\hspace{3mm}w_2=\frac{1-\alpha_2}{1+\alpha_2}
    \label{EoS_power_law}
,\end{equation}
which are both constants. However, these parameters, as opposed to the single-field case, will not generally be indicative of the scaling rate of each component with the expansion. This is due to the effective interaction taking place between both fields, which affects their cosmic evolution. On the other hand, in this case equations \eqref{rho1p_cov} and \eqref{rho1kin_cov} yield
\begin{eqnarray}
    \rho_1&=&\lambda_1V_1(\phi_1)(1-\alpha_1)Y^{\alpha_1},
    \label{rho1_power_law0} \\
    \rho_2&=&\lambda_2(1+\alpha_2)X_2Y^{\alpha_2}.
    \label{rho2_power_law0}
\end{eqnarray}
So, we will require $\alpha_2>-1$, in order not to violate the weak energy condition (WEC), and $\alpha_1<1$ so that $\rho_1$ is also positive \cite{Dario1:2023cor}.
{\color{black} Now, noting that the EoM for $\phi_1$, \eqref{EoM_phi_cov_pot}, indicates that $\phi_1=\mathrm{const.}$; and substituting \eqref{EoM_T^mu_1kin_cov} for the $\phi_2$ field, we obtain that}

\begin{eqnarray}
    \rho_1&=&\lambda_1V_1(1-\alpha_1)Y^{\alpha_1},
    \label{rho1_power_law} \\
    \rho_2&=&\frac{C_{\phi_2}}{2\lambda_2Y^{\alpha_2}a^6}(1+\alpha_2),
    \label{rho2_power_law}
\end{eqnarray}
where $V_1$ and $C_{\phi_2}=C_\phi^2$ are constants. So, in order to know the evolution of the energy densities with the scale factor, first we need to obtain $Y(a)$ from the constrain equation \eqref{EoM_T^mu_mixto_general}, as we shall show.

Lastly, we will introduce the following parameters that will be useful for the rest of the work. We will parameterize the scaling of each component and its time evolution introducing an effective EoS parameter function $w_{\mathrm{eff},i}(a)$ for each component so that the individual effective conservation equations are satisfied \cite{dpm:2024ewm}:
\begin{equation}
    \rho_i'+3\frac{a'}{a}[1+w_{\mathrm{eff},i}(a)]\rho_i=0,
    \label{indiv_cons}
\end{equation}
which allows us to obtain the following results when recalling the expressions for the energy densities \eqref{rho1_power_law} and \eqref{rho2_power_law}:
\begin{align}
    w_{\mathrm{eff},1}(a)&=-\frac{1}{3}\alpha_1a\frac{Y'(a)}{Y(a)}-1,
    \label{weff1}\\
    w_{\mathrm{eff},2}(a)&=\frac{1}{3}a\left[\frac{6}{a}+ \alpha_2\frac{Y'(a)}{Y(a)}\right]-1,
    \label{weff2}
\end{align}
which describe the scaling behavior of each component at all times and can be directly computed after solving $Y(a)$ from the constraint \eqref{EoM_T^mu_mixto_general}. 
\subsection{Single-field domination regimes}\label{secIVA}
We will start by studying the phenomenology of these types of models when one of the fields dominates the cosmic evolution over the other. Firstly, let us rewrite the constraint \eqref{EoM_T^mu_mixto_general} using the EoM for $\phi_1$ and $\phi_2$, which are equations \eqref{EoM_phi_cov_pot} and \eqref{EoM_T^mu_1kin_cov}, respectively. This allows us to generally obtain $Y$ as a function of $a$ as
\begin{equation}
    \frac{1}{2}\frac{C_{\phi_2}H_2'(Y)}{H_2(Y)^2a^6}-H_1'(Y)V_1=\mathcal{C}_g,
    \label{constraint_mixto_cov}
\end{equation}   
which in the power-law case reads
\begin{equation}
    \frac{C_{\phi_2}}{2\lambda_2}\frac{\alpha_2}{a^6Y^{\alpha_2+1}}-\lambda_1V_1\alpha_1Y^{\alpha_1-1}=\mathcal{C}_g,
    \label{constraint_mixto_power}
\end{equation}
where $\mathcal{C}_g$ is a constant. If the potential field is dominant, the main contribution in \eqref{constraint_mixto_cov} will be the second term in its l.h.s.~and $\phi_1$ will approximately depict a cosmological constant. Since $V_1$ is a constant, this would imply that 
\begin{equation}
    H_1'(Y)=\mathrm{const.},
    \label{constr_pot_dom}
\end{equation}
which, as mentioned before, translates into $Y=\mathrm{const.}$\footnote{There is also the possibility that $H_1(Y)=aY+b$, but this is just a sub-case in which $T^{(1)\mu\nu}$ would be conserved and there would not be an effective interaction.} Substituting a constant value of $Y=Y_c$ in \eqref{rho2_power_law} would then yield
\begin{equation}
    \rho_2=\frac{C_{\phi_2}(1+\alpha_2)}{2\lambda_2Y_c^{\alpha_2}a^6}\propto a^{-6},
    \label{rho2_pot_dom}
\end{equation}
which indicates that, regardless of the coupling exponent $\alpha_2$, the kinetic field $\phi_2$ will behave as a stiff fluid with $w_\mathrm{eff,2}=1$, whereas $w_\mathrm{eff,1}=-1$ under the potential field domination. This is a consequence of the interaction between the fields. It should be noted that this is the only possible behavior for a canonical Diff kinetically driven field.

\begin{figure}[H]
    \centering
    \includegraphics[scale=0.58]{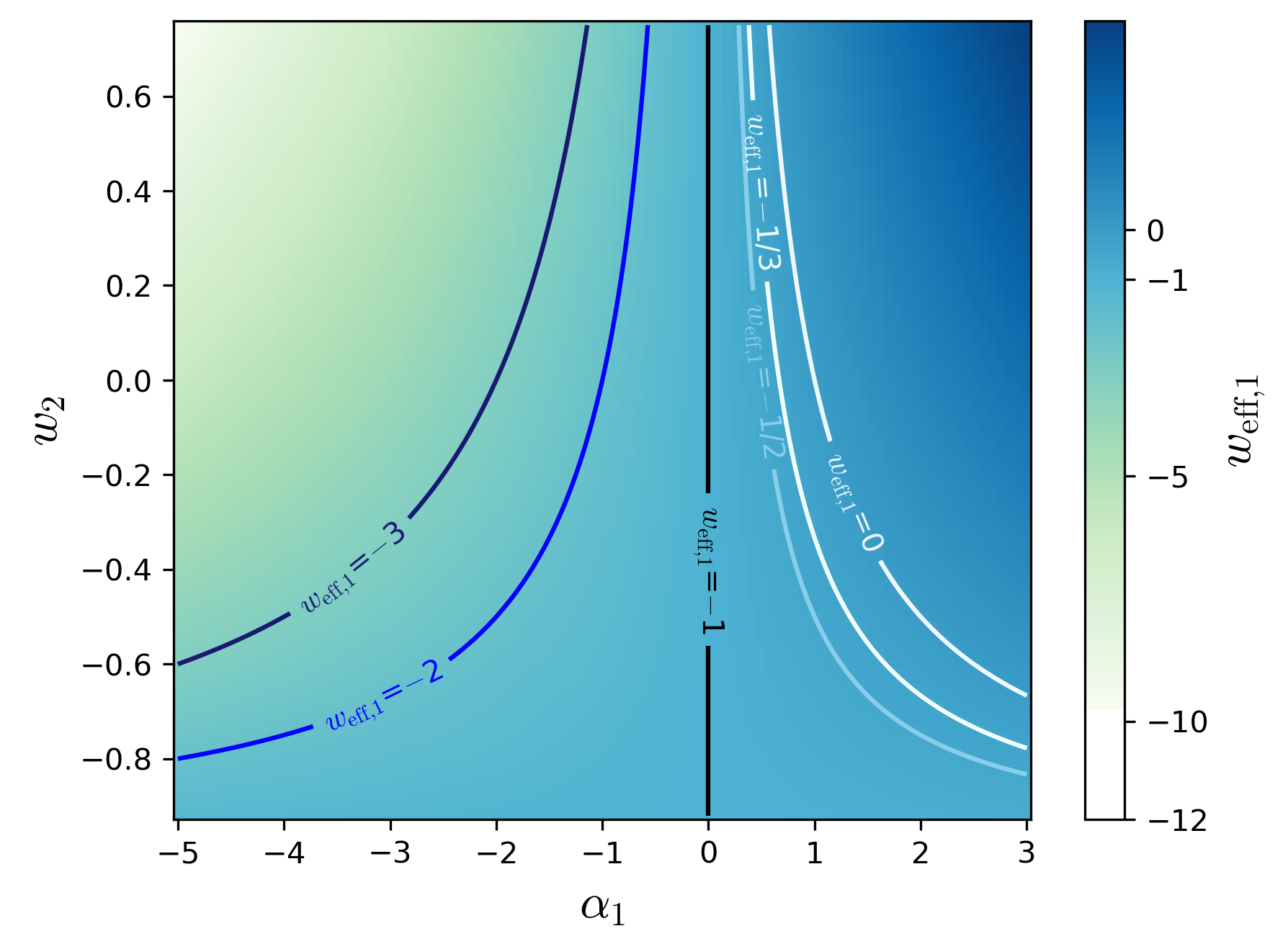}
    \caption{\justifying Effective EoS parameter $w_\mathrm{eff,1}$ for the potentially driven field under the kinetic field domination regime in terms of the potential coupling exponent $\alpha_1$ and the EoS parameter for the kinetic field $w_2$.}
    \label{Fig1}
\end{figure}

On the other hand, when the kinetic field $\phi_2$ is dominant, the primary contribution to the l.h.s. of \eqref{constraint_mixto_cov} will be its first term and $\phi_2$ will approximately behave as expected from its EoS parameter (see equation\eqref{EoS_power_law}). From equation \eqref{constraint_mixto_power}, this implies 
\begin{equation}
    Y(a)=\left(\frac{\alpha_2C_{\phi_2}}{2\lambda_2\mathcal{C}_g}\right)^{\frac{1}{1+\alpha_2}}a^{-\frac{6}{1+\alpha_2}}.
    \label{constr_kin_dom}
\end{equation}
Thus, if we substitute this result in equation \eqref{rho1_power_law} we obtain 
\begin{equation}
   \rho_1\propto a^{-\frac{6\alpha_1}{1+\alpha_2}},
    \label{rho1_kin_dom}
\end{equation}
during the {\color{black}kinetic} domination regime, so that:
\begin{equation}
    -\frac{6\alpha_1}{1+\alpha_2}=-3(1+w_\mathrm{eff,1})\Rightarrow w_\mathrm{eff,1}=\alpha_1(1+w_2)-1
    \label{weff_pot}
\end{equation}
and $w_\mathrm{eff,2}=0$. Thus, we can appreciate that there are multiple types of behavior available for the field $\phi_1$ under the kinetic domination regime. Thus, due to the interaction that results from the symmetry breaking, the energy density of the field whose dynamics is driven by its potential energy does not keep a constant value, but evolves with the cosmic expansion. In figure \ref{Fig1} all possible behaviors for $\phi_1$ under $\phi_2$ domination are summarized.

In light of this analysis, these models can provide a wide range of phenomenological possibilities for interacting scalar fields in a cosmological background. In particular, we have shown that the EoS parameters presented in \eqref{EoS_power_law} are only associated with the dynamical evolution of the fields in the regime when they are dominant, while their evolution is affected by the interaction taking place through the symmetry breaking when they are subdominant. More concretely, the kinetic field will always scale like a stiff fluid when it is subdominant, and the potential field will be able to present multiple types of behavior under kinetic domination, considerably differing from a cosmological constant. This is particularly illuminating, as it allows us to see that we can have a dynamical dark energy component purely driven by a potential field, which will always depict a cosmological constant in the asymptotic regime where it dominates, while the phenomenology can be different in the subdominant regime. As we can see in figure \ref{Fig1}, both quintessence ($-1<w_\mathrm{eff,1}<-1/3$) and phantom ($w_\mathrm{eff,1}<-1$) scaling ratios are accessible. This could be of notable interest when it comes to the study of the dark sector, as it could lead to interacting dark matter-dark energy models with a dark energy density that increases with the expansion (phantom) under dark matter domination in the past, which has been favored by the latest observations \cite{DESI:2025zgx}. 

\subsection{Energy exchange}\label{secIVB}
We will now discuss the energy exchange and the evolution of its direction for this type of models. In particular, we will consider both single-field domination regimes and study whether the interacting kernel $Q$ changes sign between them in terms of the parameters of our model. Let us start by analyzing the case in which the kinetic field $\phi_2$ is dominant, i.e., we will approximately have 
\begin{eqnarray}
    0&=&\rho_1'+3\frac{a'}{a}(\rho_1+p_1)+\rho_2'+3\frac{a'}{a}(\rho_2+p_2)\nonumber\\ 
    &\simeq&\rho_2'+3\frac{a'}{a}(\rho_2+p_2).
    \label{Q_field2_k}
\end{eqnarray}    
We can obtain the interacting kernel $Q$ by substituting in \eqref{Q_field1} the results in the kinetic domination regime, those being \eqref{rho1_power_law} and \eqref{constr_kin_dom}. This yields
\begin{equation}
    Q=-k\alpha_1(1-\alpha_1)a'a^{\frac{-1-\alpha_2-6\alpha_1}{1+\alpha_2}},
    \label{Q_kin_dom}     
\end{equation}
where $k\equiv6\lambda_1V_1/(1+\alpha_2)[\alpha_2C_{\phi_2}/(2\lambda_2 \mathcal{C}_g)]^{1/(1+\alpha_2)}$ is a positive constant factor. We can thus outline the different cases in table \ref{Tab1}, where we can {\color{black}see} that if $\alpha_1<0$, the potential field gains energy from the kinetic field, while the opposite happens if $0<\alpha_1<1$. This is particularly interesting, because it shows us that the phantom behavior for the potential field under kinetic domination (which corresponds to $\alpha_1<0$ as shown in figure \ref{Fig1}) is a consequence of the energy it receives from the kinetic field through the interaction. Finally, we would like to remind the reader that we need to have $\alpha_1<1$ for a positive $\rho_1$.
\begin{table}[h]
\begin{tabular}{|c|c|c|}
\hline
                 & $\alpha_1<0$ & $0<\alpha_1<1$  \\ \hline
$\mathrm{sign}(Q)$ & $+$          & $-$                      \\ \hline
\end{tabular}
\caption{\justifying Direction of the energy exchange in the kinetic domination regime.}
\label{Tab1}
\end{table}

On the other hand, we can proceed with a similar analysis for the case in which the potential field is dominant, in which we will approximately have that
\begin{eqnarray}
    0&=&\rho_1'+3\frac{a'}{a}(\rho_1+p_1)+\rho_2'+3\frac{a'}{a}(\rho_2+p_2)\nonumber\\ &\simeq&\rho_1'+3\frac{a'}{a}(\rho_1+p_1)
    \label{Q_field2_p}
\end{eqnarray}
and $Y(a)$ will approximately be constant. We can thus obtain the interacting kernel $Q$ by substituting in equation \eqref{Q_field2} the results in the potential domination regime, those are \eqref{rho2_power_law} and $Y=Y_c$. This yields 
\begin{equation}
    Q=\sigma \alpha_2 a'a^{-7}Y_c^{-\alpha_2}=\sigma\frac{1-w_2}{1+w_2}a'a^{-7}Y_c^{-\alpha_2},
    \label{Q_pot}
\end{equation}
where $\sigma\equiv6C_{\phi_2}/(2\lambda_2)$ and $Y_0$ are positive constants\footnote{Note that $Y(a)$ is defined to be a positive function to have a well-defined TDiff frame change.}. Taking into account that $w_2>-1$, so that the {\color{black} weak energy condition} is not violated, the results for this regime are summarized in table \ref{Tab2}. In this particular case, we see that if $-1<w_2<1$, the kinetic field will be giving energy to the potential field, whereas the contrary will happen for $w_2>1$. We can hence observe that the direction of the energy exchange will only change between both single-field domination regimes when $-1<w_2<1$ and $0<\alpha_1<1$ or when $w_2>1$ but $\alpha_1<0$.
\begin{table}[h]
\begin{tabular}{|c|c|c|}
\hline
                 & $-1<w_2<1$ & $w_2>1$ \\ \hline
$\mathrm{sign}(Q)$ & $+$        & $-$     \\ \hline
\end{tabular}
\caption{\justifying Direction of the energy exchange in the potential domination regime.}
\label{Tab2}
\end{table}

\subsection{Dark sector phenomenology}\label{secIVC}
The results presented up to this point can be specially relevant when it comes to the study of the dark sector. Indeed, as we previously discussed when presenting the asymptotic single-field domination regimes, $\phi_2$ will tend to its typical scaling (proportional to $a^{-3(1+w_2)}$) under kinetic domination, and $\phi_1$ will asymptotically approach a cosmological constant behavior under potential domination. This opens up a wide range of phenomenological scenarios that could be useful for describing an interacting dark sector. In particular, fixing $\alpha_2=1$, we will have $w_2=0$ and $\phi_2$ will asymptotically behave as cold dark matter. 
There are two interesting cases for such dark sector models. The first one corresponds to $\alpha_1<0$, which is associated with phantom DE behavior of $\phi_1$ in the past, according to the previously presented single-field domination regimes and figure \ref{Fig1}. In this case, as we can see in table \ref{Tab1} and table \ref{Tab2}, $\phi_1$ would gain energy from $\phi_2$, allowing it to depict phantom behavior in the past ($w_\mathrm{eff,1}<-1$). This energy exchange, however, will not be as strong in the future, where $\phi_1$ will tend to an asymptotic cosmological constant behavior from the phantom region and $\phi_2$ would scale as a stiff fluid. In the second case, on the other hand, we have that $0<\alpha_1<1$. As we can see in figure \ref{Fig1}, this range of values for $\alpha_1$ leads to quintessential dark energy behavior under kinetic domination for the $\phi_1$ component, i.e., $-1<w_\mathrm{eff,1}<-1/3$. In this case, we can infer from tables \ref{Tab1} and \ref{Tab2} that $\phi_2$ would gain energy from $\phi_1$ at early times, so $\phi_1$ cannot present phantom behavior when its contribution is subdominant; however, $\phi_1$ would gain energy from $\phi_2$ when it dominates the dynamics.

We emphasize that the interaction is reflected through the particular shape of the $Y(a)$ function obtained from solving the constraint \eqref{constraint_mixto_power}, which can generally only be done numerically. Considering our particular dark sector scenario ($\alpha_2=1$) and dividing \eqref{constraint_mixto_power} by $\mathcal{C}_g$ we can obtain a more compact version of this equation:
\begin{equation}
    \frac{C_2}{a^6 Y^2}-\alpha_1C_1Y^{\alpha_1-1}=1,
    \label{constr_compacto}
\end{equation}
where $C_2\equiv C_{\phi_2}/(2\lambda_2\mathcal{C}_g)$ and $C_1\equiv \lambda_1 V_1/\mathcal{C}_g$. From this expression, one could argue that there might be an ambiguity when calculating $Y(a)$, as the \textit{initial condition} $Y_0$ at the current time depends on two arbitrary constants, $C_1$ and $C_2$. In fact, we would have that, at the current time ($a_0=1$):
\begin{equation}
    C_2=Y_0^2+\alpha_1C_1Y_0^{\alpha_1+1}.
    \label{condition_Y0}
\end{equation}
However, it is not difficult to check whether this arbitrariness has physical implications or not. In fact, since Einstein equations are preserved, we can compute the Hubble rate from the zeroth-component $G^0{}_0=8\pi GT^0{}_0$, which is no other than Friedmann equation:
\begin{equation}
    \frac{\Dot{a}^2}{a^2}=H^2=\frac{8\pi G}{3}(\rho_b+\rho_1+\rho_2),
    \label{Friedmann_1}
\end{equation}
where $\rho_b$ denotes the usual baryonic density, which in this study is considered to be a free and non-interacting Diff component, while $\rho_1$ and $\rho_2$ depict the interacting TDiff dark sector. We also denoted $(\cdot)\equiv\dd/\dd t$, where we performed the time reparametrization $\dd t=b(\tau)\dd\tau$ to work using cosmic time. It is worth mentioning that, since this study focuses on researching the dark sector and its background properties at the matter and dark energy epochs, we neglected radiation for the sake of simplicity. Thus, if we take into account the cosmic sum rule ($\rho_1^0+\rho_2^0=(1-\Omega_b)\rho_\mathrm{c}$, where $\Omega_b$ is the baryonic abundance and $\rho_\mathrm{c}$ the critical density) and introduce the following parameter relating the $\phi_2$ and $\phi_1$ abundances today
\begin{equation}
    \gamma\equiv\frac{\rho_2^0}{\rho_1^0}=\frac{2C_2}{C_1(1-\alpha_1)Y_0^{\alpha_1+1}},
    \label{gamma_parameter}
\end{equation}
we can obtain that
\begin{align}
    \Omega_1\equiv \frac{\rho_1^0}{\rho_\mathrm{c}}=\frac{1-\Omega_b}{1+\gamma},
    \label{Omega_1}\\
    \Omega_2\equiv \frac{\rho_2^0}{\rho_\mathrm{c}}=\gamma\frac{1-\Omega_b}{1+\gamma},
    \label{Omega_2}
\end{align}
Since both components are interacting with each other, they can only be interpreted as DE and DM asymptotically.
Therefore, the Hubble rate will read
\begin{equation}
    \begin{split}
    H^2=H_0^2\left[\Omega_ba^{-3}+\Omega_1\left(\frac{Y}{Y_0}\right)^{\alpha_1}+\right.\left.\Omega_2a^{-6}\left(\frac{Y_0}{Y}\right)\right].
    \end{split}
    \label{Hubble_parameter}
\end{equation}This only depends on the quotient $\tilde{Y}\equiv Y/Y_0$, and we can thus conclude that the physics of the model will not depend on $Y_0$, which will simply be a normalization constant for $Y$ and can be absorbed in the definition of $\gamma$. Consequently, in order to simplify the expressions, we will set the condition $Y_0=1$ for the rest of the work\footnote{The freedom of choice for the initial condition of the additional field $Y_0$ translates into an initial condition in the no longer gauge extra degree of freedom $b_0$ when working in the TDiff approach (check appendix \ref{appendix} for further understanding of the correspondences between the two approaches).}.

Let us also now introduce the total dark sector equation of state for these models, which will later be discussed for each case and can be directly computed from the total conservation law
\begin{equation}
    \rho_\mathrm{tot}'(a)+ \frac{3}{a}[\rho_\mathrm{tot}(a)+p_\mathrm{tot}(a)]=0,
    \label{total_cons_law}
\end{equation}
which translates into 
\begin{equation}
    w_\mathrm{DS}(a)=-\frac{a\rho_\mathrm{tot}'(a)}{3\rho_\mathrm{tot}}-1,
    \label{DS_EoS}
\end{equation}
where $\rho_\mathrm{tot}(a)=\rho_1(a)+\rho_2(a)$. Once expressed in terms of the parameters of our model, \eqref{DS_EoS} will read
\begin{equation}
    \begin{split}
    w_\mathrm{DS}(a)=-\frac{a}{3}\frac{\Omega_1\alpha_1Y'Y^{\alpha_1-1}}{\Omega_1Y^{\alpha_1}+\frac{1}{Ya^6}\Omega_2} 
    +\frac{\Omega_2a^{-6}\left(-\frac{6}{aY}-\frac{Y'}{Y^2}\right)}{\Omega_1Y^{\alpha_1}+\frac{1}{Ya^6}\Omega_2}-1.
    \end{split}
    \label{DS_EoS_final}
\end{equation}
We will now separately analyze the $\alpha_1<0$ and the $0<\alpha_1<1$ cases.
\subsubsection{Phantom case $(\alpha_1<0)$}
\begin{figure}[h]
    \centering
    \includegraphics[scale=0.54]{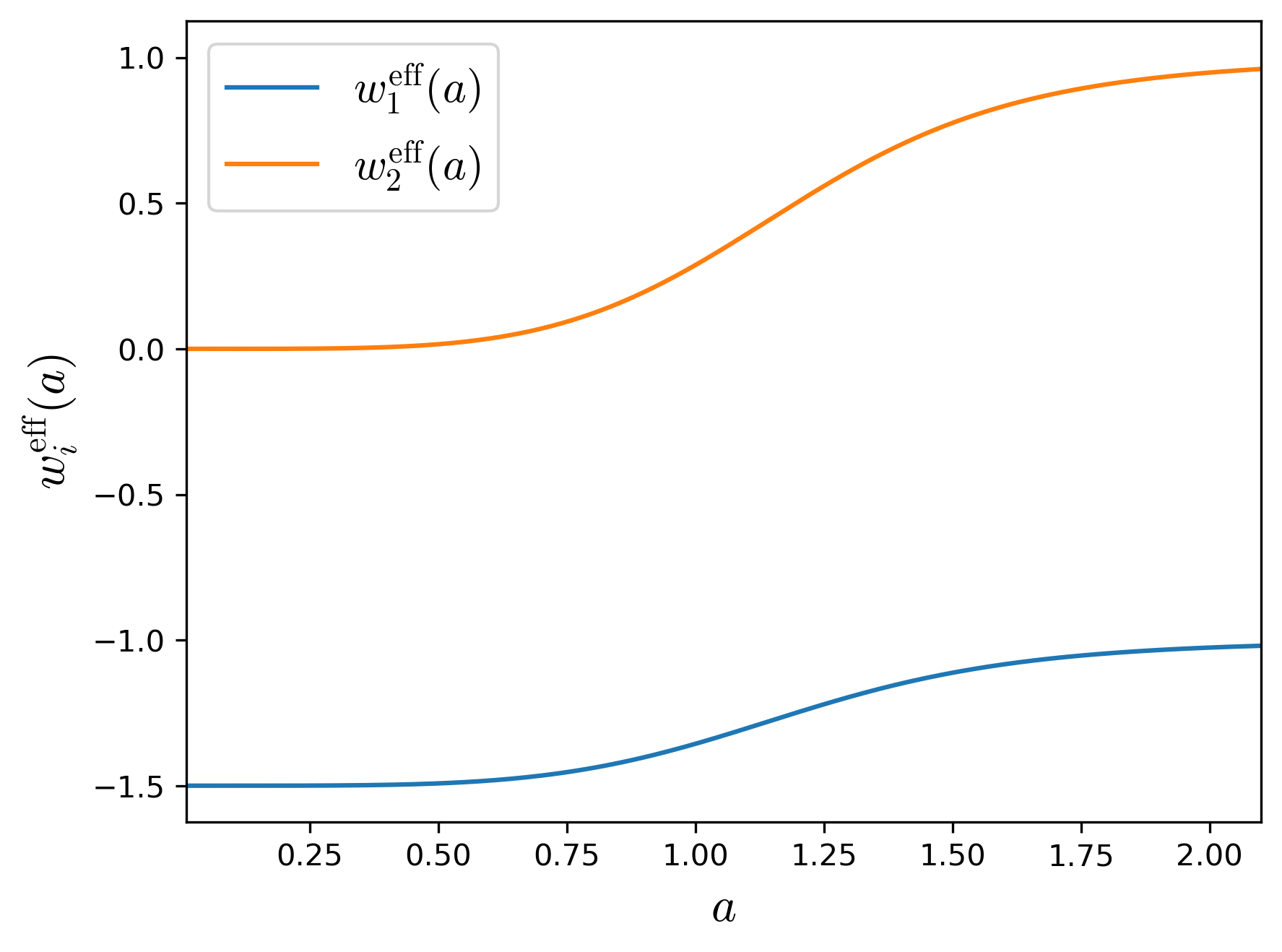}
    \caption{\justifying Effective EoS parameter for both components in terms of the scale factor for the case with $\alpha_1<0$ ($\alpha_1=-1/2$, $C_1=7/10$). Dark energy exhibits phantom behavior and asymptotically tends to a cosmological constant in the future.}
    \label{Fig2}
\end{figure}

\begin{figure}[t]
    \centering
    \includegraphics[scale=0.54]{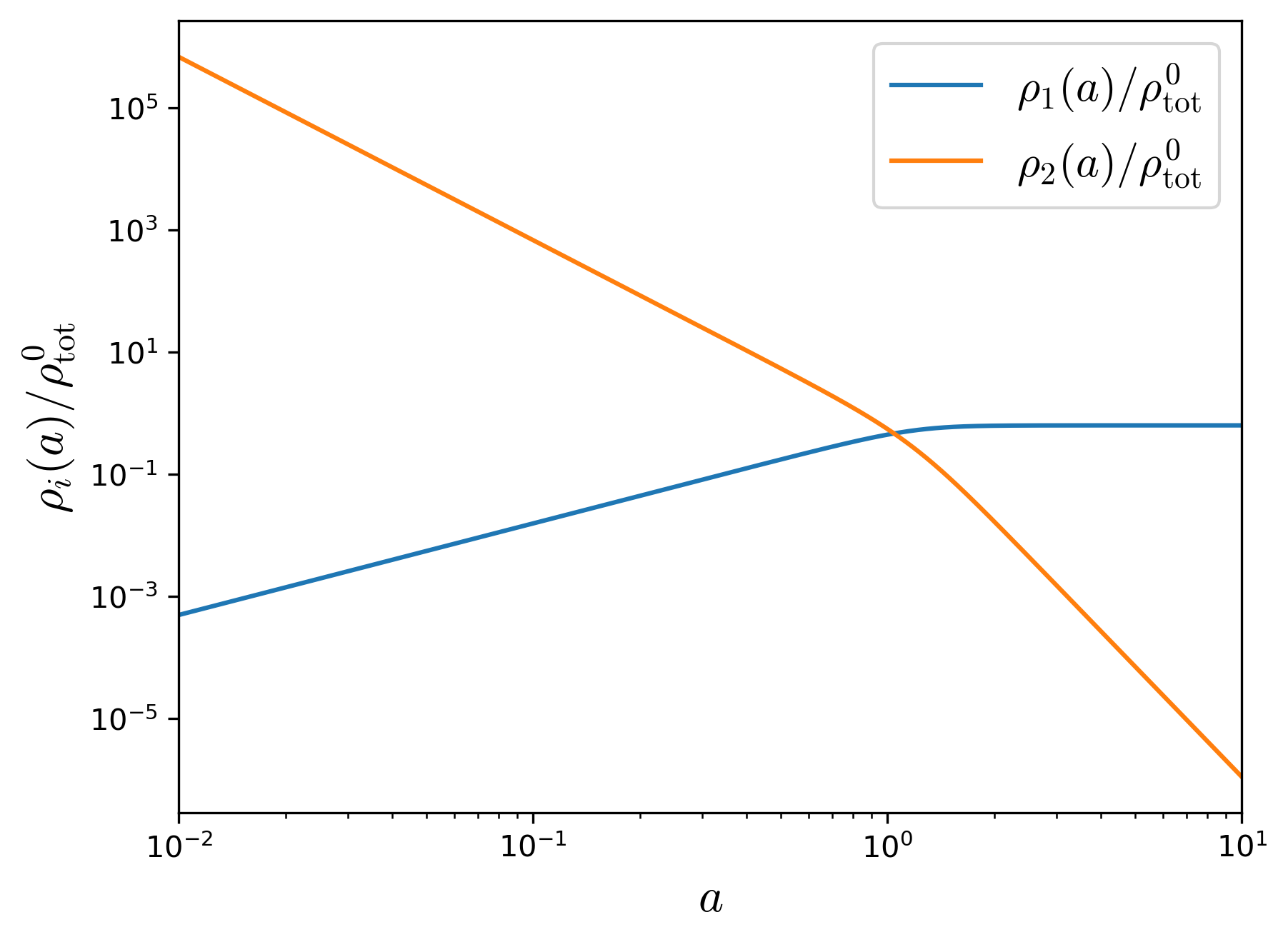}
    \caption{\justifying Evolution of the DM and DE energy densities in terms of the scale factor $a$ for $\alpha_1<0$ ($\alpha_1=-1/2$, $C_1=7/10$). DE presents phantom behavior and asymptotically tends to a cosmological constant in the future.}
    \label{Fig4}
\end{figure}

\begin{figure}[b]
    \centering
    \includegraphics[scale=0.54]{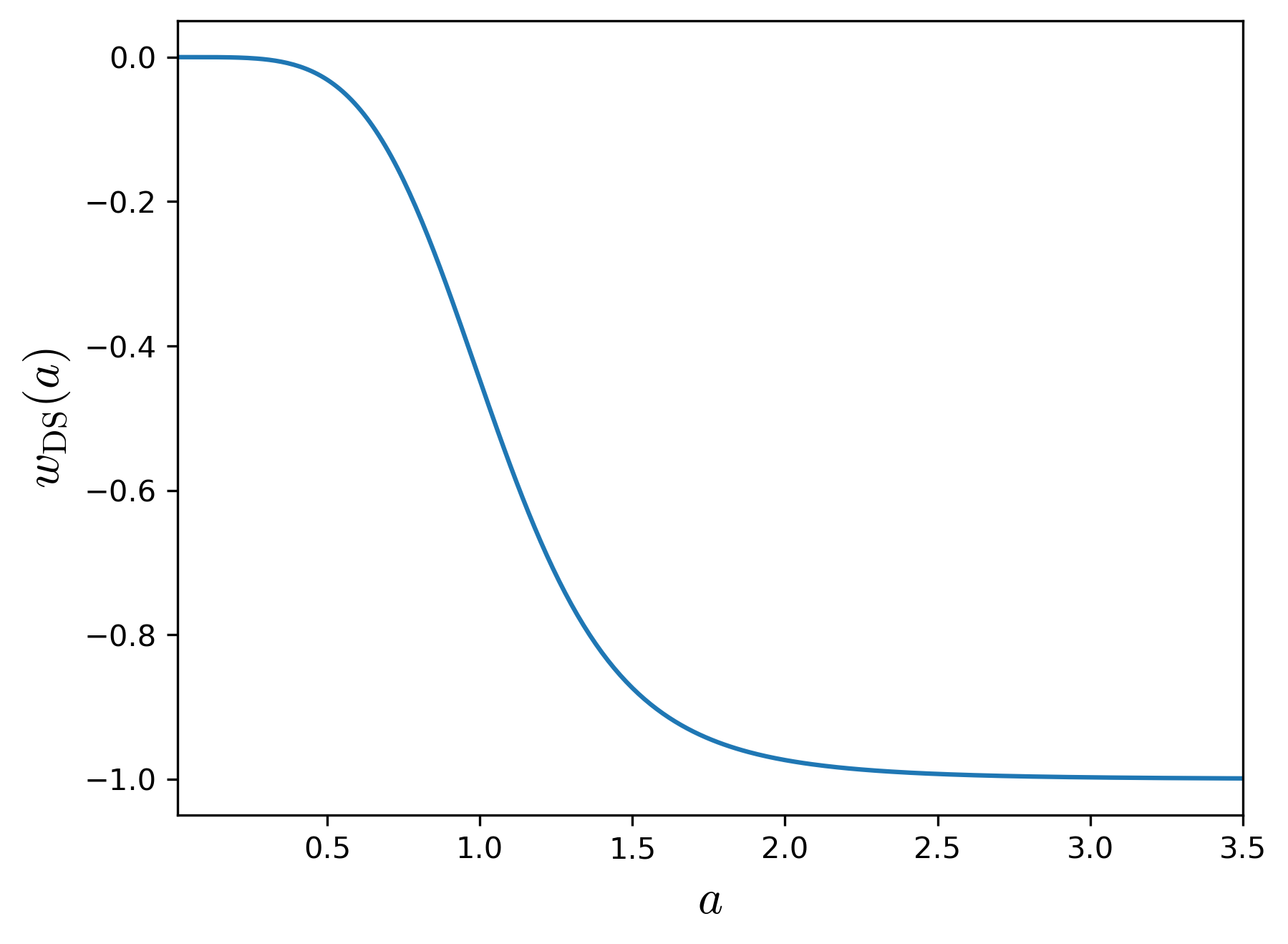}
    \caption{\justifying Equation of state of the dark sector in terms of the scale factor $a$ for $\alpha_1<0$ ($\alpha_1=-1/2$, $C_1=7/10$). $w_\mathrm{DS} $ smoothly transitions from 0 in the past (DM domination) to $-1$ in the future (DE domination).}
    \label{Fig5}
\end{figure}

If $\alpha_1<0$, as we previously discussed, the DE component will present phantom behavior ($w_\mathrm{eff,1}<-1$) as a consequence from it receiving energy from DM. This energy exchange also remains in the same direction when DE becomes dominant, but it will not be as strong and $\phi_1$ will asymptotically tend to a cosmological constant. The effective behavior of each component and its evolution with the expansion is represented in figure \ref{Fig2}. Similarly, we can also see this when looking at figure \ref{Fig4}, in which the evolution of the energy densities is represented in terms of the scale factor.
The evolution of the effective dark sector EoS parameter is also represented in figure \ref{Fig5}, where we can observe that the combined dark sector in this case presents a type of solutions that reproduce a typical behavior similar to that of $\Lambda$CDM, with the total dark sector EoS transitioning smoothly from zero to minus one. This is interesting because it implies that, even if we have a phantom component, $\rho_\mathrm{tot}+p_\mathrm{tot}>0$ and the effective dark sector behavior will not be phantom. This allows us to avoid the singularities that typically appear in phantom dark energy models \cite{Nojiri:2005sx}.
However, even if we analyzed the phenomenology behind this model and the evolution of the energy exchange, we have not discussed why it is allowed for $\phi_1$ to become dominant in the future and reach its asymptotic behavior. The explanation to this arises when looking at the constraint, i.e., the EoM for the extra field \eqref{constr_compacto}. In particular, we could parameterize the leading order of $Y(a)$ as a power-law in $a$ and from now on we will focus on the study of its asymptotic behavior when $a\rightarrow\infty$. With regards to this, taking the derivative of the constraint \eqref{constr_compacto} allows us to obtain $Y'(a)$ the following way
\begin{equation}
    Y'(a)=-\frac{6C_2Y(a)}{a(2C_2-a^6C_1\alpha_1Y(a)^{1+\alpha_1}+a^6C_1\alpha_1^2Y(a)^{1+\alpha_1})},
    \label{Y_prima}
\end{equation}
which shows us that, since $C_{1,2}>0$ and $\alpha_1<1$, $Y'(a)<0$ and $Y$ will always be a decreasing function, allowing us to parameterize its asymptotic leading order as $Y(a)\simeq ka^{-\beta}$, with $\beta\geq0$ and $k$ a constant. Introducing this in the constraint \eqref{constr_compacto} yields
\begin{equation}
    \frac{1+\alpha_1C_1}{k^2a^6}a^{2\beta}-\alpha_1C_1k^{\alpha_1-1}a^{\beta-\beta\alpha_1}=1.
    \label{constr_power_asymptotic_gen}
\end{equation}
When $\alpha_1<0$, we can rewrite the asymptotic constraint as
\begin{equation}
    \frac{(1+\alpha_1C_1)}{k^2}a^{2\beta-6}-\alpha_1C_1k^{\alpha_1-1}a^{\beta|\alpha_1|+\beta}=1,
    \label{constr_phantom}
\end{equation}
which only has a solution for $a\rightarrow\infty$ when $\beta=0$ and thus $Y\rightarrow k=\mathrm{const.}$ in the asymptotic future, which, when substituted in \eqref{constr_phantom} implies that 
\begin{equation}
    k=\left(-\frac{1}{\alpha_1C_1}\right)^\frac{1}{\alpha_1-1}.
    \label{k_phantom}
\end{equation}
Notice that this is also consistent with the fact that for $\alpha_1<0$ DE gets to dominate in the future, since introducing $\beta=0 $ results in the contribution from $\phi_2$ in the constraint \eqref{constr_compacto} becoming negligible when $a\rightarrow\infty$. Similarly, we can also see from \eqref{Y_prima} that $Y'(a)$ will {\color{black}tend to zero} in the future. Thus, recalling the expressions \eqref{weff1} and \eqref{weff2} for the effective EoS parameters of each component and introducing \eqref{Y_prima} in them thus yields:
\begin{align}
w_{\mathrm{eff},1}(a)=-1+\frac{2C_2\alpha_1}{2C_2+a^6 C_1(\alpha_1-1)\alpha_1Y(a)^{1+\alpha_1}},
\label{weff_1_y_prima} \\
w_{\mathrm{eff},2}(a)=\frac{C_1a^6(\alpha_1-1)\alpha_1 Y(a)^{1+\alpha_1}}{2C_2+a^6 C_1(\alpha_1-1)\alpha_1 Y(a)^{1+\alpha_1}}.
\label{weff_2_y_prima}
\end{align}
Therefore, we can tell that, since $Y\rightarrow\mathrm{const.}$ in the future and thus $aY'/Y\rightarrow0$ in this case (see \eqref{Y_prima}), $w_{\mathrm{eff},1}\rightarrow-1$ and $w_{\mathrm{eff},2}\rightarrow 1$ as expected from the numerical results presented previously. This result is particularly interesting, as it allows us to analytically see that for $\alpha_1<0$ DE will be able to reach its asymptotic domination regime. 

\subsubsection{Tracking case $(0<\alpha_1<1)$}
\begin{figure}[t]
    \centering
    \includegraphics[scale=0.54]{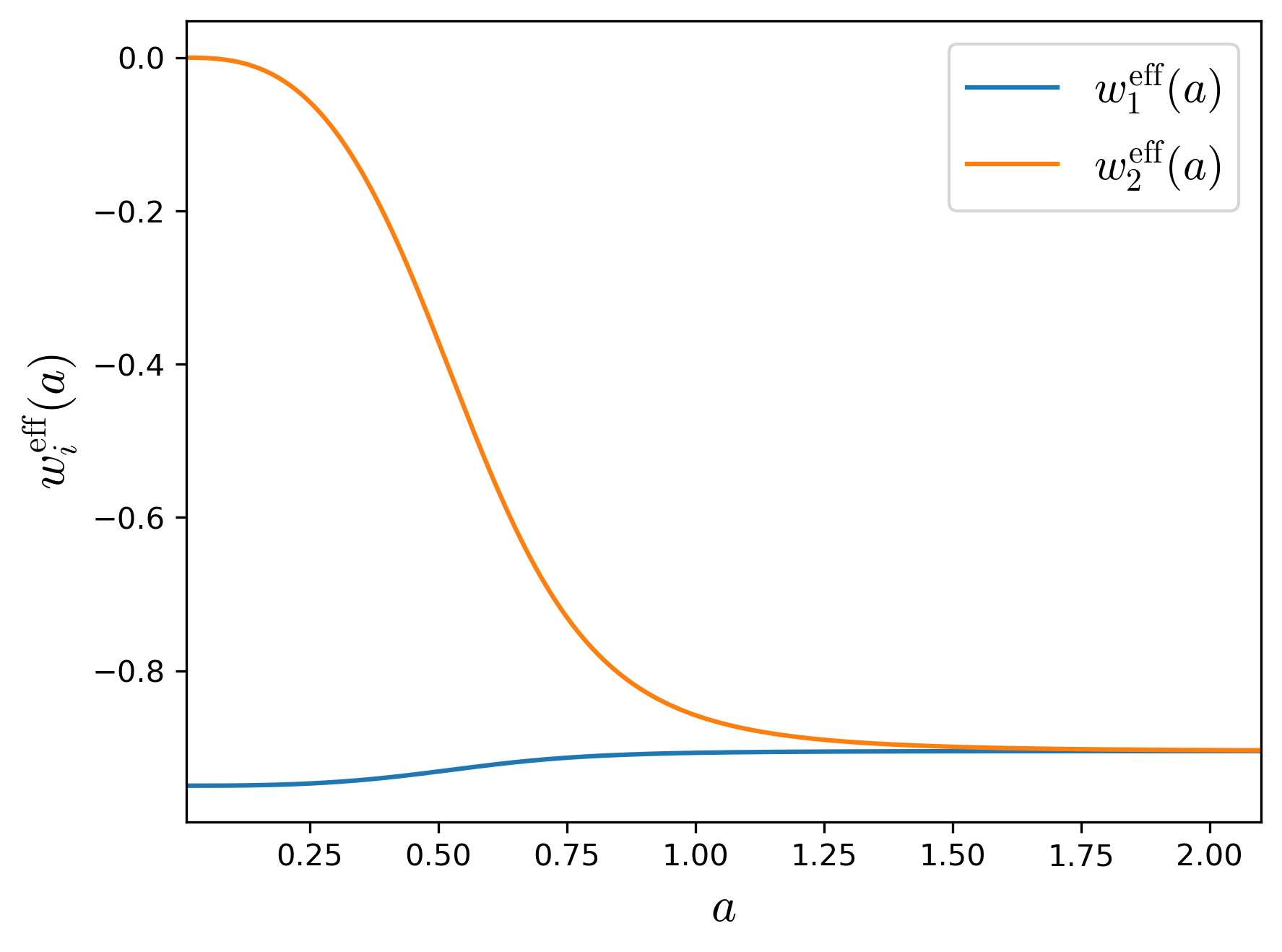}
    \caption{\justifying Effective EoS parameter for both components in terms of the scale factor for the case with $0<\alpha_1<1$ ($\alpha_1=1/20$, $C_1=700$). $\phi_1$ exhibits phantom quintessence behavior with $-1<w_{\mathrm{eff},1}(a)<-1/3$ under $\phi_2$ domination in the past and tracks $\phi_2$ in the future.}
    \label{Fig3}
\end{figure}

\begin{figure}[b]
    \centering
    \includegraphics[scale=0.54]{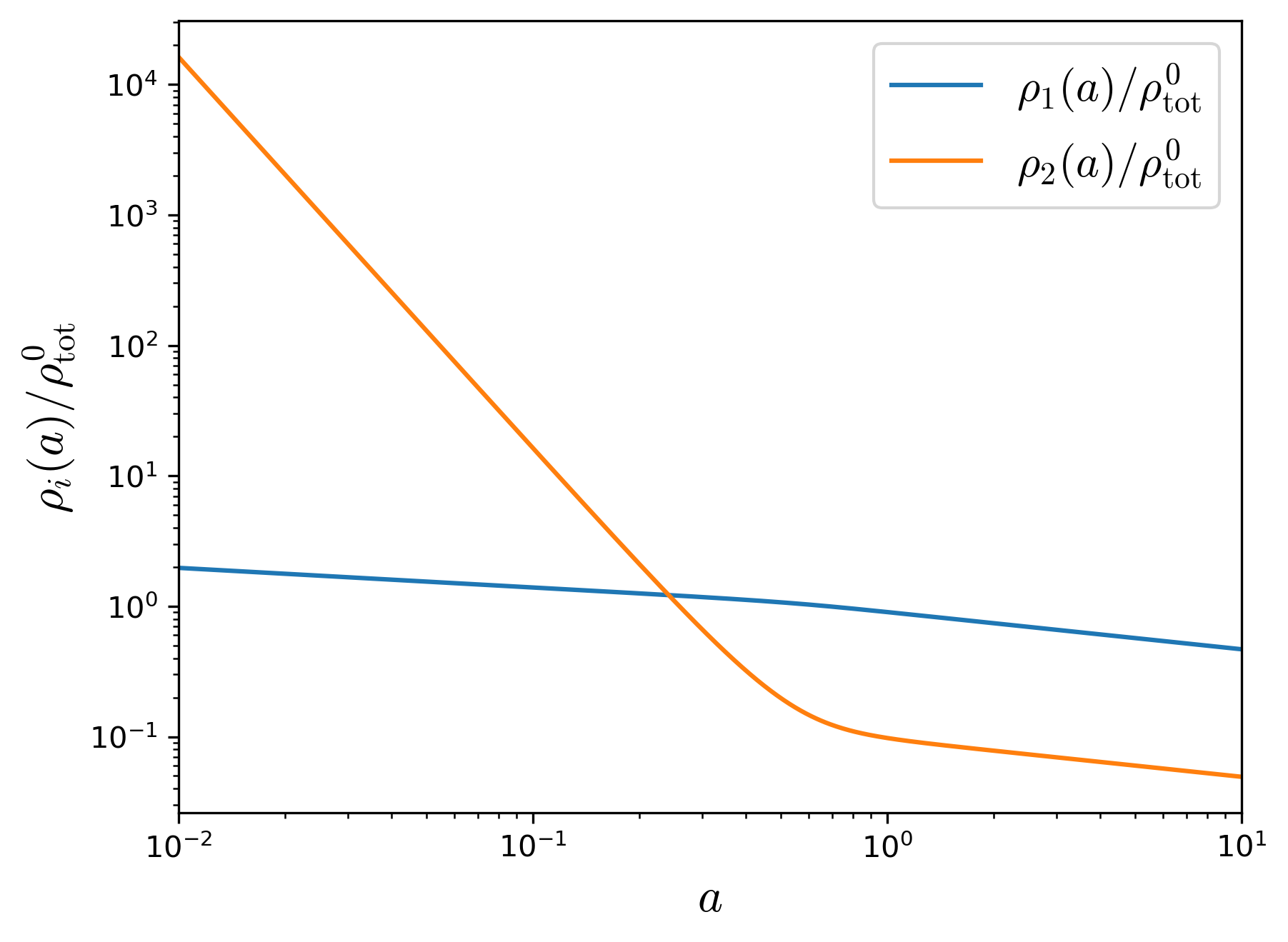}
    \caption{\justifying Evolution of the $\phi_2$ and $\phi_1$ energy densities in terms of the scale factor $a$ for $0<\alpha_1<1$ ($\alpha_1=1/20$, $C_1=700$). $\phi_1$ decays with the expansion, and both components track each other in the future.}
    \label{Fig6}
\end{figure}

As it was previously discussed, for $0<\alpha_1<1$, the $\phi_1$ component (asymptotic DE) will present a typical quintessence scaling under $\phi_2$ (asymptotic DM) domination, as it loses energy in favor of $\phi_2$. However, as we can see in figure \ref{Fig3} where we plotted the evolution of the individual EoS parameters with the expansion, $\phi_1$ does not reach the expected behavior in which it depicts a cosmological constant. In fact, both $\phi_2$ and $\phi_1$ start tracking each other in the future, presenting the same {\color{black}behavior} (notice that this makes the notion of dark matter and dark energy more diffuse, since both components would behave in the same way as a consequence of the interactions).  In a similar way, we can also appreciate this phenomenology in terms of the energy densities in figure \ref{Fig6}. As we shall discuss, this will be a consequence of $\phi_1$ not being able to become sufficiently dominant when $0<\alpha_1<1$\footnote{This is a consequence of $\phi_1$ losing energy in this scenario, as we previously discussed, which results in both fields slowly starting to track each other in the future.}.
More concretely, rewriting \eqref{gamma_parameter} allows us to understand why it is not possible for the potential field $\phi_1$ to be dominant enough. In fact, recalling \eqref{condition_Y0} and taking into account that $Y_0=1$, we will have:
\begin{equation}
    \gamma=\frac{2}{C_1(1-\alpha_1)}+\frac{2\alpha_1}{1-\alpha_1}.
    \label{gamma_simplificado}
\end{equation}
 We can see that, if $C_1\ll 1$, $\gamma\gg1$ and $\phi_1$ will be much less abundant than $\phi_2$ today, making it impossible for it to dominate. Similarly, if $C_1\sim 1$ we can see from \eqref{gamma_simplificado} that $\gamma\gtrsim1$, which also makes it impossible for $\phi_1$ to be dominant at the current time, with the energy exchange stopping this from happening and resulting in both fields tracking each other and DE not becoming dominant enough to present its asymptotic cosmological constant behavior. However, if $C_1\gg1$ it is not difficult to see that $\gamma\rightarrow2\alpha_1/(1-\alpha_1)$, which indicates that the only possibility for $\phi_1$ to dominate today would occur when $\alpha_1\lesssim1/5$ and $C_1\gg1$ or $C_1\rightarrow\infty$, which is not what we are looking for, since it is equivalent to enforcing $\phi_1$ domination at all times. As we will later show, despite there being some cases allowing for $\phi_1$ to be dominant today, this dominance will not be enough for it to display a cosmological constant behavior and the tracking will always happen.

\begin{figure}[h]
    \centering
    \includegraphics[scale=0.54]{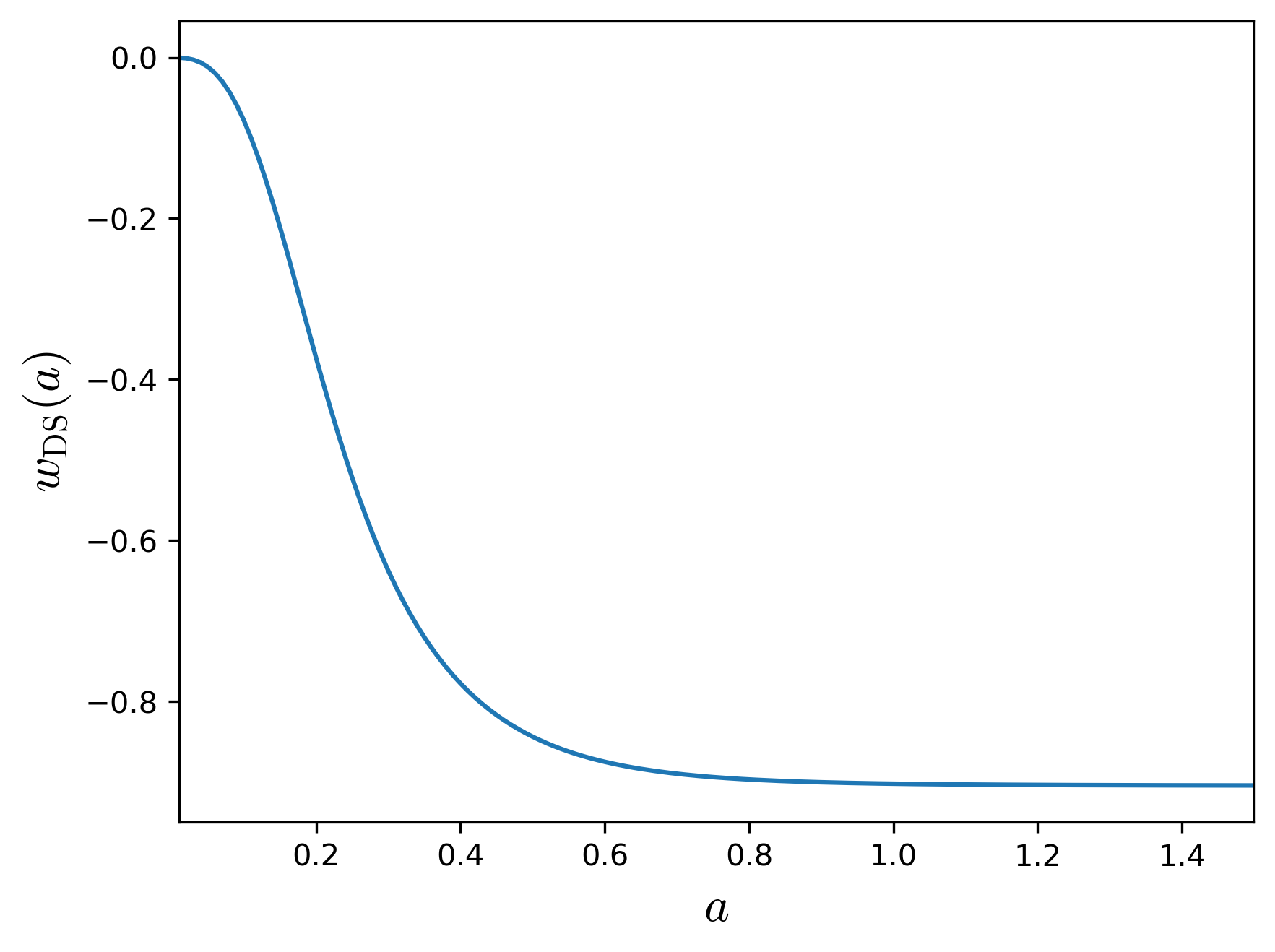}
    \caption{\justifying Equation of state of the dark sector in terms of the scale factor $a$ for $0<\alpha_1<1/2$ ($\alpha_1=1/20$, $C_1=700$). $w_\mathrm{DS}$ asymptotically approaches $w_\mathrm{track}$ in \eqref{wtracking}.}
    \label{Fig7}
\end{figure}

Thus, in order to understand the nature of the tracking in this case, we can perform an asymptotic analysis similar to the one in the previous case. More specifically, we can rewrite the constraint at the leading order $(Y(a)\simeq ka^{-\beta})$ {\color{black}when $a\rightarrow\infty$}:
\begin{equation}
    \frac{(1+\alpha_1C_1)}{k^2a^6}a^{2\beta}-\alpha_1C_1k^{\alpha_1-1}\frac{a^{\beta}}{a^{\beta\alpha_1}}=1.
    \label{asymptotic_constr_no_phantom}
\end{equation}
The only way for the l.h.s. of \eqref{asymptotic_constr_no_phantom} to be finite occurs when $|\beta|>3$ considering that both terms have the same dependence with the scale factor $a$, which imposes the following constraint on $\beta$:
\begin{equation}
    2\beta-6=\beta(1-\alpha_1)\implies\beta=\frac{6}{1+\alpha_1}.
    \label{beta_condition}
\end{equation}
However, this would give us an indetermination in \eqref{asymptotic_constr_no_phantom}, which can only vanish if the constants multiplying each term are equal, i.e.,
\begin{equation}
    k=\left(\frac{1+\alpha_1C_1}{\alpha_1C_1}\right)^{\frac{1}{1+\alpha_1}}.
    \label{k_relation_no_phantom}
\end{equation}
If we substitute $Y(a)=ka^\frac{-6}{1+\alpha_1}$ in the effective EoS parameters \eqref{weff1} and \eqref{weff2}, we will obtain that, in the asymptotic future:
\begin{equation}
    w_{\mathrm{eff},1}=w_{\mathrm{eff},2}= w_\mathrm{track}\equiv\frac{\alpha_1-1}{\alpha_1+1},
    \label{wtracking}
\end{equation}
which explains the tracking phenomena we previously discussed (this can also be checked from \eqref{weff_1_y_prima} and \eqref{weff_2_y_prima}). Therefore, in light of \eqref{wtracking}, we can see that the tracking value of the EoS parameter of each component, which is also equivalent to the asymptotic EoS parameter of the dark sector, will be such that $-1<w_\mathrm{track}<0$ for $0<\alpha_1<1$. This will thus allow the universe in our model to have an asymptotic accelerated expansion once the tracking regime is reached for $\alpha_1<1/2$, and the moment when this occurs will depend on the rest of the parameters of the model, namely $\gamma$. It is also worth mentioning that, as opposed to the $\alpha_1<0$ case, we will now have non-negligible contributions from both fields in the future (see \eqref{rho1_power_law} and \eqref{rho2_power_law}). Finally, for the sake of completion we include in figure \ref{Fig7} 
the evolution of the dark sector EoS parameter with the expansion for a particular model with $0<\alpha_1<1/2$such that there is an accelerated expansion at the current time.

Lastly, we would also like to discuss the choice of $\phi_1$ to be dominated by its potential. As we stated at the beginning of this section, we are considering that $\phi_1$ is purely dominated by its potential term during the matter and dark energy epochs, which may seem as a strong requirement on the model. However, as we saw from \eqref{Y_prima}, $Y(a)$ will always be a decreasing function in terms of the scale factor, so it would be enough to assume that the coupling function of the kinetic term of $\phi_1$, i.e., $H_{K1}(Y)$ in \eqref{S_mixto_gen_cov} decreases faster than $H_{V_1}(Y)$ during this time period. It should be emphasized that no requirements on the kinetic term of $\phi_1$ are forced before this time period, we only require potential domination for the matter and dark energy epochs, in a similar way as the slow-roll approximation in inflation. 
\section{Conclusions}\label{secV}
In this work, we have considered TDiff invariant theories in the matter sector, consisting of two free scalar fields, {\color{black}one being dominated by its potential ($\phi_1$) and the other by its kinetic term ($\phi_2$)}, and studied their cosmological dynamics at the background level. Since the Diff symmetry is only broken through the matter sector of the action, the Bianchi identities are preserved and the conservation-law for the total EMT is satisfied. However, this conservation is no longer trivial {\color{black}and introduces a physical constraint}. {\color{black}Breaking the symmetry thus results in having less gauge freedom to fix the geometry. In particular, when working in a cosmological background, we can no longer fix the time coordinate to one by means of TDiff transformations;} and the conservation law for the EMT results in a constraint involving both matter fields that allows us to obtain the no longer gauge degree of freedom (the lapse function) in terms of the scale factor. This constraint is different to the one obtained in the single-field case, and it entails an energy exchange between both fields that can be interpreted as an effective interaction, even if there is no interacting potential introduced in the Lagrangian.

In light of this, we also presented the covariantized formalism. In this formalism, we introduce an additional field whose EoM will be equivalent to the constraint derived from the EMT conservation in the TDiff approach. We chose to work with this formalism instead of doing so with the TDiff approach since it is more convenient to use and it makes the physics of the model easier to unveil. In particular, we have considered power-law coupling functions of the additional field for each component ($H_i(Y)\propto Y^{\alpha_i}$) and have studied the phenomenology in a cosmological context. We first studied the single-field domination regimes and observed that the kinetic field will always depict a stiff fluid behavior under potential domination. The potential field will be able to present a wide range of possible scalings under kinetic domination as a consequence of the interactions, including phantom or quintessence dark energy behavior. With this in mind, we put this into practice by studying the particular applications of this model with regards to the dark sector, by assuming $\alpha_2=1$ so that $\phi_2$ depicts DM and leaving $\alpha_1$ free for the dark energy component. We showed that, for $\alpha_1<0$, DE will always gain energy from DM, which results in DE being able to present phantom behavior, asymptotically tending to a cosmological constant behavior (as expected from a potentially driven field). On the other hand, for $0<\alpha_1<1$, the potential field will lose energy in favor of the kinetic one, which results in the EoS parameter for $\phi_1$ being such that $-1<w_\mathrm{eff,1}<-1/3$ under DM domination in the past. This results in $\phi_1$ not being able to become sufficiently dominant in the future, making it so that both components start tracking each other in the future, with $w_\mathrm{track}<-1/3$ for $0<\alpha_1<1/2$. Nevertheless, both scenarios are phenomenologically interesting and could present a possible dynamical dark energy model, as we also saw that both could be compatible with having an accelerated late-time expansion. 

Finally, we would like to remark that this work provides further insight of multi-field TDiff models and their cosmological consequences at the background level. We analyzed the phenomenological possibilities and explored the effective interactions taking place between the fields as a consequence of breaking the Diff symmetry from a theoretical perspective, discussing their relevance when it comes to describing a dynamical dark sector. However, for a deeper and more complete understanding of TDiff theories, future projects include developing the perturbation theory for TDiff models and performing a full observational analysis and a comparison of the theory with cosmological observables and data.

\section*{Acknowledgements}
The authors would like to thank Darío
Jaramillo-Garrido for useful comments and suggestions. DTB acknowledges financial support from Universidad Complutense de Madrid and Banco Santander
through the Grant No. CT25-24.
This work has been supported by the MICIN (Spain)
Project No. PID2022-138263NB-I00 funded by MICIU/AEI/10.13039/501100011033 and by ERDF/EU.

\appendix
\section{TDiff vs Covariantized approach}\label{appendix}
In this appendix we will briefly discuss the covariantized approach and we will explore the relationship {\color{black} with the TDiff formalism} in a deeper way. As we saw in section III, \eqref{EoM_T^mu_mixto_general} is equivalent to \eqref{constr_mixto_general} in the TDiff frame. However, working with \eqref{constr_mixto_general} is much more complicated, as it involves solving an ODE to obtain $b(a)$, while \eqref{EoM_T^mu_mixto_general} is just an {\color{black} algebraic equation} that allows us to obtain $Y(a)$. Not only is it easier to solve and obtain $Y(a)$ from \eqref{EoM_T^mu_mixto_general}, but, in turn, it is also straightforward to recover $b(a)$ from this considering the TDiff limit, since $Y\mapsto 1/(ba^3)$. 

Nevertheless, if we wanted to work using the TDiff approach, we could solve \eqref{constr_mixto_general} and obtain $b(a)$. More concretely, if we recall Friedmann equation \eqref{Friedmann_1} we can see that 
\begin{equation}
    \frac{\dd}{\dd\tau}=\sqrt{\frac{8\pi G}{3}}(ab)\rho^{1/2}\frac{\dd}{\dd a},
    \label{Friedmann_subs}
\end{equation}
which allows us to obtain:
\begin{equation}
    b'(a)=-\frac{b(a)^{2\hat{\alpha}_1-1}(6\hat{\alpha}_1-3)a^{6\hat{\alpha}_1-4}}{a^{6\hat{\alpha}_1-3}(2\hat{\alpha}_1-1)b(a)^{2\hat{\alpha}_1-2}+\frac{4C_2}{C_1\hat{\alpha}_1} a^{-3}},
    \label{ODE_b(a)}
\end{equation}             
where $\hat{\alpha}_1\equiv(1-\alpha_1)/2$, since, as we discussed in section \ref{secIII}, power-law functions of $Y$ translate into power-law functions of $g$ when performing the TDiff limit, but with a different exponent, i.e. \cite{Dario2:2024tdv},
\begin{equation}
    Y^{\alpha_1}\mapsto g^{\hat{\alpha}_1}, \hspace{3mm}\hat{\alpha}_1=\frac{1}{2}(1-\alpha_1).
    \label{translation_power_law}
\end{equation}
Thus, we see that solving \eqref{ODE_b(a)} will allow us to obtain {\color{black} $b(a)$} in the TDiff approach, with there being an arbitrariness in the initial condition $b_0=b(a_0)$. However, in the covariantized approach we simply had to solve an equation (see \eqref{constraint_mixto_cov}) to obtain $Y(a)$ and then perform the TDiff limit to obtain $b(a)$ in a much more straightforward way than in the TDiff case. In this case, as we discussed in section \ref{secIV}, the arbitrariness comes through the dependence of the model on the integration constants stemming from  the EoM of the fields, which is reflected in $Y_0$ and thus in $b_0$ in the TDiff frame. In the TDiff frame, on the other hand, this arbitrariness comes directly as an integration constant that appears when solving the ODE for $b(a)$.

As we showed in section \ref{secIV}, there is no physical dependence on the arbitrary parameter $Y_0$, as it does not appear in the physical quantities. This must indicate that, in a similar manner, $b_0$ must not be physically meaningful for our model when working with the TDiff approach. In fact, if we set $a_0=1$,  we will see that $Y_0\mapsto1/b_0$ in the TDiff frame. Therefore, there is a correspondence between the TDiff and covariantized approaches. If we work using the covariantized approach, there will be an arbitrariness on the initial condition of the additional field $Y_0$, which translates into an initial condition on {\color{black}$b_0$} in the TDiff frame. Thus, if we recall the expressions for the energy densities in the TDiff formalism \eqref{rho_1pot} and \eqref{rho1kin} and use the cosmic sum rule, we can see that
\begin{equation}
    \gamma=\frac{\rho_2^0}{\rho_1^0}=\frac{2C_2}{C_1(1-\alpha_1)} b_0^{2-2\hat{\alpha}_1},
    \label{gamma_TDiff}
\end{equation}
which is actually equivalent to \eqref{gamma_parameter} after taking the TDiff limit and allows us to obtain for the Hubble rate (similarly to how we did in \eqref{Hubble_parameter}):
\begin{eqnarray}
    H^2=&H_0^2\left[\Omega_ba^{-3}+\Omega_1\hat{b}(a)^{2\hat{\alpha}_1-1}a^{6\alpha_1-3}+\right.\nonumber\left.  \Omega_\mathrm{2}\hat{b}(a)a^{-3}\right],
    \label{H(z)_TDiff}
\end{eqnarray}
where we defined
\begin{equation}
    \Omega_\mathrm{2}\equiv\gamma\frac{1-\Omega_b}{\gamma+b_0^{2\hat{\alpha}_1-2}};
    \label{ODM_TDiff}
\end{equation}
and where $\hat{b}(a)\equiv b(a)/b_0$, and therefore the model does not depend on $b_0$, since $\hat{b}_0=1$ and $\hat{b}(a)$ will not depend on $b_0$ (this can be checked in a straightforward way when performing the variable change $\dd\hat{b}=\dd b/b_0$ in \eqref{ODE_b(a)}). This is in accordance with the results discussed in section \ref{secIV} concerning the covariantized approach. However, even if the TDiff approach is more complicated to work with, it is also very useful. More specifically, if we {\color{black}go to the TDiff frame} in \eqref{Hubble_parameter} we will recover \eqref{H(z)_TDiff}, which presents a more explicit dependence on the physical parameters (the scale factor and the shift function), and could be of valuable use when performing a deeper analysis of the model. On the other hand, the fact of having to solve an ODE in the TDiff approach makes it no longer trivial to obtain a relationship between $b_0$, $C_1$ and $C_2$ (similar to \eqref{condition_Y0}), and it will not be as straightforward to understand the tracking phenomena that take place for $0<\alpha_1<1$ ($0<\hat{\alpha}_1<1/2$).

\bibliographystyle{elsarticle-num} 
\bibliography{bibliografia} 
 \end{document}